\documentclass[letterpaper,titlepage,11pt]{article}

\pdfoutput=1

\usepackage{amsmath,amssymb,amsthm,mathrsfs,bbm}
\usepackage{latexsym,amscd,amsbsy,amsfonts,dsfont}
\usepackage{graphicx}
\usepackage[
      colorlinks=true,
      linkcolor=blue,
      urlcolor=blue,
      filecolor=black,
      citecolor=red,
      pdfstartview=FitV,
      pdftitle={},
        pdfauthor={Tomas Andrade, Roberto Emparan, David Licht, Raimon Luna},
        pdfsubject={},
        pdfkeywords={},
        pdfpagemode=None,
        bookmarksopen=true
      ]{hyperref}\usepackage{caption}

\newcommand{\sR}{\mathsf{R}}

\newcommand{\beq}{\begin{equation}}
\newcommand{\eeq}{\end{equation}}
\newcommand{\beqa}{\begin{eqnarray}}
\newcommand{\eeqa}{\end{eqnarray}}
\newcommand{\bea}{\begin{eqnarray}}
\newcommand{\eea}{\end{eqnarray}}

\newcommand{\nn}{\nonumber}

\newcommand{\ie}{{i.e.,\,}}
\newcommand{\eg}{{e.g.,\,}}

\newcommand{\lp}{\left(}
\newcommand{\rp}{\right)}

\newcommand{\ord}[1]{{\mathcal O}\lp #1\rp}

\def\clock{{\count0=\time
           \divide\count0 60
           \ifnum\count0<10 0\fi\the\count0
           \multiply\count0 -60 \advance\count0 \time
           :\ifnum\count0<10 0\fi \the\count0
         }}
\newcommand{\timestamp}{{\small\vbox{\hbox{\tt\jobname.tex}
\hbox{\the\day/\the\month/\the\year, \clock}}}}

\usepackage{bm}
\usepackage{latexsym}
\usepackage[latin1]{inputenc}
\usepackage{amsfonts}
\usepackage{amssymb}
\usepackage{amsmath}
\usepackage{subfigure}

\numberwithin{equation}{section}

\begin{document}

\begin{titlepage}
\leftline{}
\vskip 2cm
\centerline{\LARGE \bf Black hole collisions, instabilities, and}
\bigskip
\centerline{\LARGE \bf cosmic censorship violation at large $D$} 
\vskip 1.6cm
\centerline{\bf Tom{\'a}s Andrade$^{a}$, Roberto Emparan$^{a,b}$, David Licht$^{a}$, Raimon Luna$^{a}$}
\vskip 0.5cm
\centerline{\sl $^{a}$Departament de F{\'\i}sica Qu\`antica i Astrof\'{\i}sica, Institut de
Ci\`encies del Cosmos,}
\centerline{\sl  Universitat de
Barcelona, Mart\'{\i} i Franqu\`es 1, E-08028 Barcelona, Spain}
\smallskip
\centerline{\sl $^{b}$Instituci\'o Catalana de Recerca i Estudis
Avan\c cats (ICREA)}
\centerline{\sl Passeig Llu\'{\i}s Companys 23, E-08010 Barcelona, Spain}
\smallskip
\vskip 0.5cm
\centerline{\small\tt tandrade@icc.ub.edu,\, emparan@ub.edu,} 
\smallskip
\centerline{\small\tt david.licht@icc.ub.edu,\, raimonluna@icc.ub.edu}

\vskip 1.cm
\centerline{\bf Abstract} \vskip 0.2cm \noindent

\noindent We study the evolution of black hole collisions and ultraspinning black hole instabilities in higher dimensions. These processes can be efficiently solved numerically in an effective theory in the limit of large number of dimensions $D$. We present evidence that they lead to violations of cosmic censorship. The post-merger evolution of the collision of two black holes with total angular momentum above a certain value is governed by the properties of a  resonance-like intermediate state: a long-lived, rotating black bar, which pinches off towards a naked singularity due to an instability akin to that of black strings. We compute the radiative loss of spin for a rotating bar using the quadrupole formula at finite $D$, and argue that at large enough $D$---very likely for $D\gtrsim 8$, but possibly down to $D=6$---the spin-down is too inefficient to quench this instability. We also study the instabilities of ultraspinning black holes by solving numerically the time evolution of axisymmetric and non-axisymmetric perturbations. We demonstrate the development of transient black rings in the former case, and of multi-pronged horizons in the latter, which then proceed to pinch and, arguably, fragment into smaller black holes. 

\end{titlepage}
\pagestyle{empty}
\small

\addtocontents{toc}{\protect\setcounter{tocdepth}{2}}

\tableofcontents
\normalsize
\newpage
\pagestyle{plain}
\setcounter{page}{1}

\section{Introduction}

\begin{figure}[h]
\centering
\includegraphics[width=1 \linewidth]{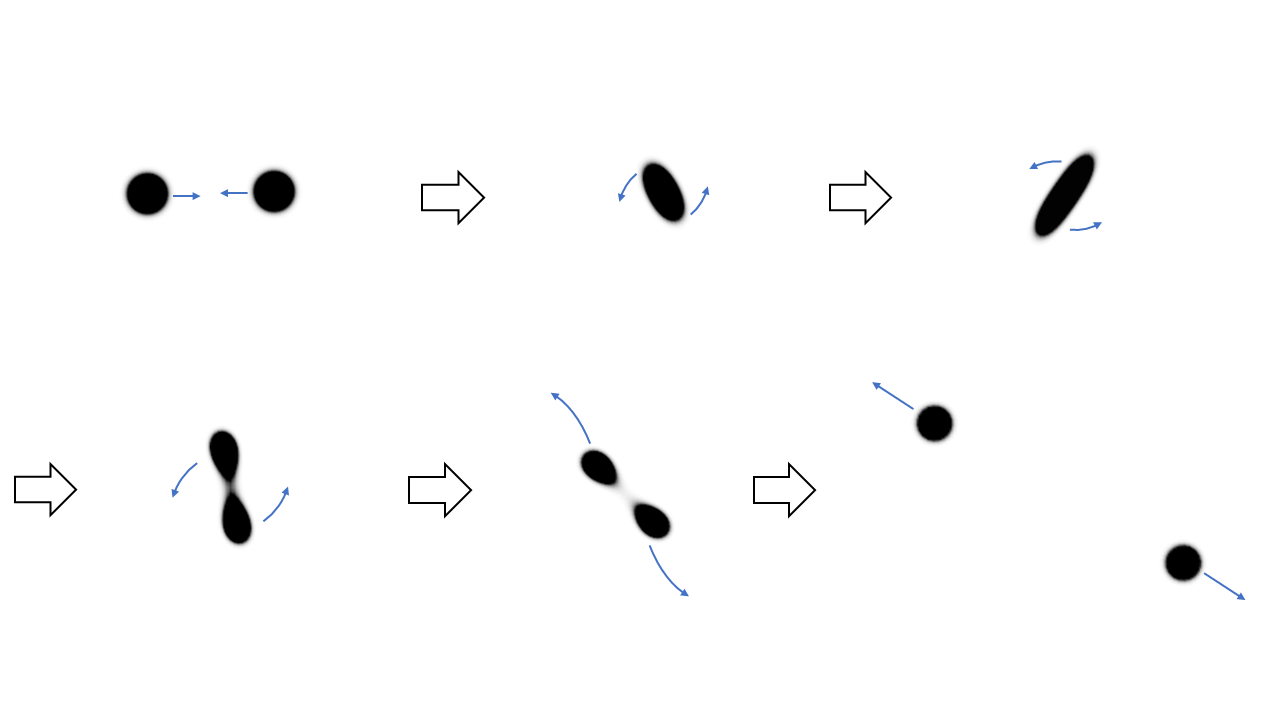}
\caption{\small Two spinning black holes collide and form a rotating black bar, which then breaks up into two outgoing black holes different than the initial ones (the figures are high-contrast density plots of the mass density obtained from the numerical simulation of a collision in the large-$D$ effective theory).}%
\label{fig:wholething}
\end{figure}

Recently \cite{Andrade:2018yqu} we have employed an effective theory of black holes in the limit of a large number of spacetime dimensions, $D\to \infty$, previously developed in \cite{Andrade:2018nsz}, in order to efficiently demonstrate processes in which two black holes collide, merge, and then---for sufficiently large total angular momentum---evolve to create a naked singularity, thus violating Cosmic Censorship (CC) \cite{Penrose:1969pc}. Afterwards, we have argued, the system plausibly `evaporates' the singularity, leaving two separate black holes that fly away from each other. The process is illustrated in figure~\ref{fig:wholething}.

The purpose of the present article is twofold: first, to provide a more detailed analysis of black hole collisions using the large-$D$ effective theory, including initially spinning Myers-Perry (MP) black holes \cite{Myers:1986un}, and elaborating on a number of  issues relevant to the violation of CC which were only cursorily discussed in \cite{Andrade:2018yqu}. Second, to extend our previous results on large-$D$ black hole dynamics by considering the time evolution of the instabilities of black holes and black bars with large spins. Ultraspinning black hole instabilities were first identified long ago \cite{Emparan:2003sy}, but black bars are a more recent addition to the panoply of higher-dimensional black hole dynamics \cite{Shibata:2010wz,Andrade:2018nsz}. They are elongated rotating black holes which can exist as stationary objects when $D\to\infty$, and which are expected to be long-lived (quasi-stationary) at large but finite $D$, since the gravitational emission from the rotating bar is suppressed like $\sim D^{-D}$.

In \cite{Andrade:2018yqu} we argued that these black bars play an important role as intermediate unstable, resonance-like states in higher-dimensional black hole collisions, and in this article we will investigate this in detail. We will show that black bars do form in these collisions, and are dynamically unstable in a manner qualitatively and even quantitatively similar to the Gregory-Laflamme (GL) instability of a black string \cite{Gregory:1993vy,Lehner:2010pn}. This is the mechanism that drives the system to the violation of CC in the black hole collision.

As we will discuss below, the presence of an intermediate quasi-stationary bar is cleanest when the total angular momentum in the merger is dominated by the initial intrinsic spin of the black holes, rather than the orbital angular momentum in the collision. In that case, the intermediate state can be closely matched to a stationary rotating black bar during several rotation cycles. In contrast, in collisions where the orbital angular momentum dominates (through sizable impact parameter or collision velocities), the intermediate state resembles less closely a stationary bar, and more a dumbbell, which pinches down more quickly than a black bar. We may think of the dumbbell as a bar at a later stage of pinching, so we regard all these collisions and decays as proceeding within the same qualitative dynamics.

Our evolutions are performed in the limit $D\to\infty$ where gravitational radiation is completely absent. One may wonder whether at finite $D$ the radiative spin-down of the bar can avert the development of the GL-like instability. To investigate this, we estimate the radiation rate using the quadrupole formula in $D$ dimensions (the emission rate of energy was obtained in \cite{Cardoso:2002pa}, while the emission rate of angular momentum is presented here for the first time). We find that the radiation is suppressed not only by high dimensionality; also the spin-down rate is small for long black bars with high spin, since their rotation velocity is slow. As a consequence, it must be possible to violate CC in a collision of two black holes if large enough total angular momentum is achieved in the intermediate, merged phase, to form a long enough black bar. 
In the terminology of \cite{Emparan:2003sy}, these results mean that at high spins `death by fragmentation' can occur more quickly than `death by radiation'. However, we have not managed to get reliable estimates for the minimum dimension in which such long, high-spin bars can form in a black hole collision. This is due to the current uncertainties in the values of the capture impact parameter and, more importantly, of the initial emission of radiation in the collision. Therefore, while CC violation is certainly possible in collisions at large enough $D$, our estimates do not allow to be equally sure about the outcome at relatively low $D$ (\eg $D=6$ or $7$).  

\begin{figure}[h]
\centering
\includegraphics[width=.3 \linewidth]{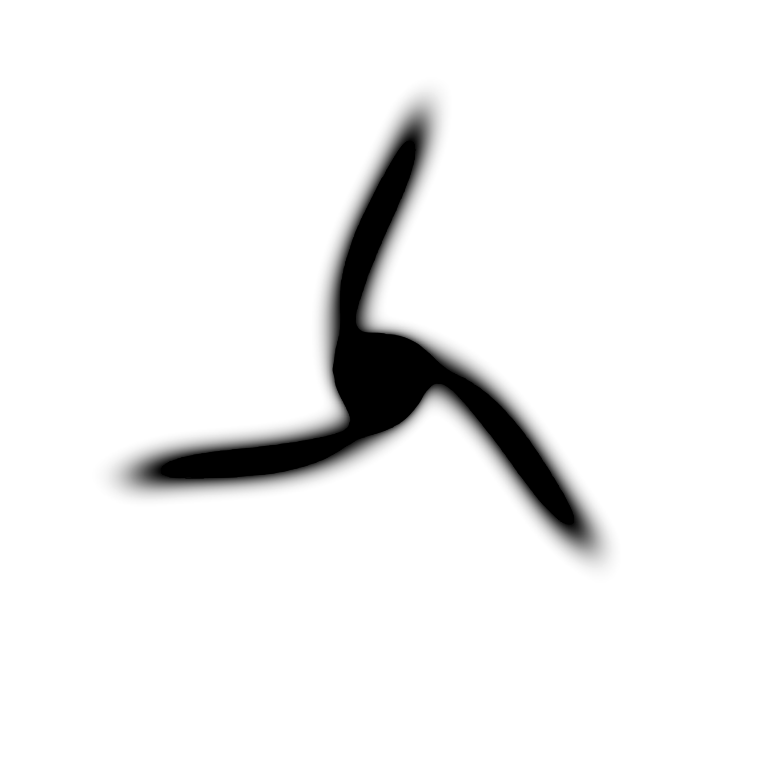}\qquad\includegraphics[width=.3 \linewidth]{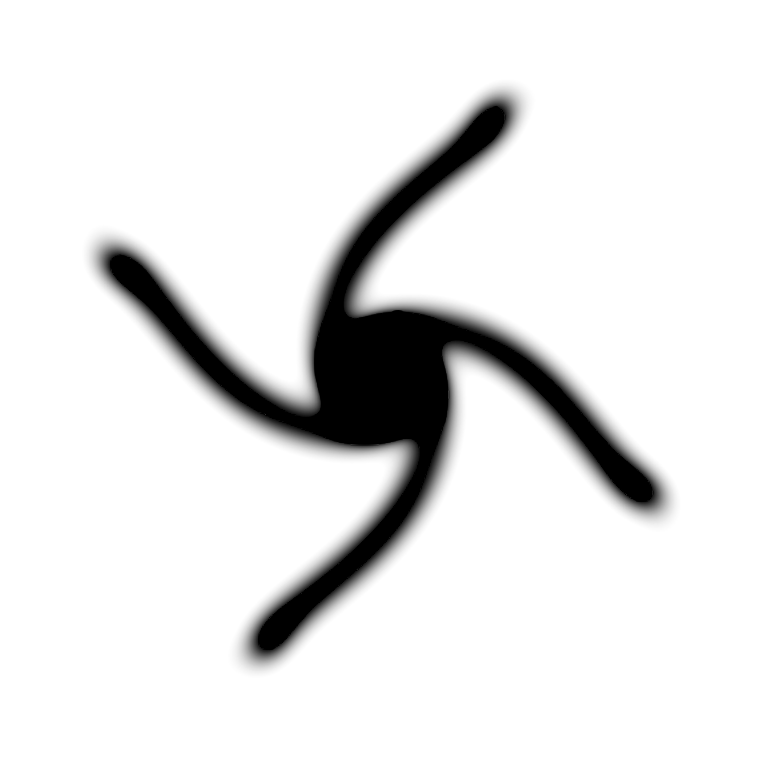}
\caption{\small Late-time horizon shape of unstable ultraspinning Myers-Perry black holes when perturbed with a tripolar and a quadrupolar mode, and then evolved with the large-$D$ effective theory. Further evolution of the black hole suggests that the arms pinch, violating CC through `death by fragmentation' \cite{Emparan:2003sy}. The quadrupolar `star' and its evolution to singular pinches has been recently observed in numerical simulations in $D=6, 7$ in \cite{Bantilan:2019bvf}.}%
\label{fig:MPstars}
\end{figure}

In addition, we present results for the non-linear evolution of instabilities of ultraspinning black holes. These exhibit remarkably rich structures, see \eg figure~\ref{fig:MPstars}, which are strikingly similar to the shapes recently observed in numerical simulations in $D=6, 7$ in \cite{Bantilan:2019bvf}. As mentioned in that article, the dynamics of these configurations is expected to lead to novel violations of CC.

\medskip

The outline of the article is the following. The next section describes the framework that we work in. Except for the final subsection, this is a summary of results in \cite{Andrade:2018nsz}.
Section \ref{sec:bhcolls} contains a detailed description of the ideas, methods, and results in our simulations of black hole collisions, including a final discussion of the violation of CC. In section \ref{sec:MP_evol_inst} we perform non-linear evolutions of the ultraspinning instabilities of MP black holes, which appear to lead to novel violations of CC. In section \ref{sec:BB_evol_inst} we study the GL-like instabilities of black bars. Our numerical results are in agreement with a model that approximates the black bar as a segment of black string. Section \ref{sec:grav_rad} estimates the radiative spin-down of black bars at finite $D$, and compares its timescale with that of the GL-like instability that drives the evolution towards CC violation. We find that the spin-down of sufficiently long bars is very inefficient in all dimensions where they exist ($D\geq 6$, and possibly $D=5$), and also for shorter bars in large enough $D$ ($D\gtrsim 8$ or possibly lower). We recapitulate and discuss the outlook of our study in section \ref{sec:discussion}. Out of the technical appendices, we remark that in appendix~\ref{app:grav radn} we compute the quadrupolar angular momentum emission rate and prove that, for bodies rigidly rotating with angular velocity $\Omega$, the emission rates of energy $E$ and angular momentum $\mathcal{J}$ satisfy
\beq
\frac{d E}{dt} =\Omega\, \frac{d\mathcal{J}}{dt}
\eeq
in all $D$.\footnote{This result was quoted, without derivation, in \cite{Emparan:2003sy}.}

\section{The effective theory and its black hole solutions}

Our main tool are the effective equations for the
dynamics of neutral, asymptotically flat black $p$-branes in the large $D$ limit. Although black branes are extended objects, we showed in \cite{Andrade:2018nsz} that they can support `blobs' in their worldvolume which capture, with remarkable accuracy and simplicity, the physics of localized black holes (of Schwarzschild, Myers-Perry, and Kerr-Newman variety \cite{Andrade:2018rcx}) in the limit $D\to\infty$, including their quasinormal spectra. In this article we will apply the approach to the study of non-linear phenomena of black holes: collisions and the long-term evolution of their instabilities.

\subsection{Effective equations}

Ref.~\cite{Emparan:2015gva} showed that one can obtain solutions of the Einstein vacuum equations in the limit $D\to\infty$ by solving a simpler set of $p+1$ partial differential equations, namely,
\begin{align}
\label{eq brane1}
	\partial_t m - \nabla^2 m &= - \nabla_i p^i\,, \\
\label{eq brane2}
	\partial_t p_i - \nabla^2 p_i &= \nabla_i m - \nabla_j \left( \frac{p_i p^j}{m} \right)\,,
\end{align}
for the $p+1$ functions $m(t,\mathbf{x} )$, $p_i (t,\mathbf{x} )$. The $\mathbf{x}=(x^i)$ denote spatial coordinates, and the indices $i,j=1,\dots,p$ are raised and lowered with the flat spatial metric $\delta_{ij}$.

The large-$D$ metric that results is
\beqa\label{largeD metric}
	ds^2 &=& 2 dt d \rho - \lp1 - \frac{m(t,\mathbf{x} )}{\sR}\rp dt^2 - \frac{2}{n} \frac{p_i(t,\mathbf{x} )}{\sR} d x^i dt \nn \\
	&+& 
	\frac{1}{n} \lp \delta_{ij} + \frac{1}{n} \frac{p_i(t,\mathbf{x} ) p_j(t,\mathbf{x} )}{m(t,\mathbf{x} ) \sR}\rp d x^i x^j + \rho^2 d \Omega_{n+1}
\eeqa
(where $n=D-p-3$). This is interpreted as a fluctuating black $p$-brane in Eddington-Finkelstein coordinates, with $\rho$ the radial coordinate orthogonal to the $p$-brane (and $\sR = \rho^n$) and $x^i$ the worldvolume spatial directions. It is apparent that $m$ measures the local horizon radius, or alternatively the local mass density \cite{Emparan:2016sjk}, while $p_i$ is associated to motion along $x^i$. More precisely, we can introduce the velocity field $v_i(t,\mathbf{x} )$ via
\begin{equation}
	p_i = \nabla_i m + m v_i\,.
\end{equation}
In terms of it, the effective equations take the hydrodynamical form of mass and momentum conservation \cite{Emparan:2016sjk}
\begin{equation}
	\partial_t m + \nabla_i( m v^i) = 0\,, \qquad \partial_t(m v^i) + \nabla_j \tau^{ij} = 0\,,
\end{equation}
with effective stress tensor
\begin{equation}
	\tau_{ij} = m \lp v_i v_j - \delta_{ij} - 2 \nabla_{(i} v_{j)} - \nabla_i \nabla_j \ln m\rp\,.
\end{equation}

\subsection{Effective theory magnitudes}
\label{sec:eff_quant}

For localized configurations, the total mass and the total linear and angular momenta, defined in terms of the effective theory variables as
\begin{align}
	M &= \int d^p x\, m(t,\mathbf{x} )\,, \\
	P_i &= \int d^p x\, p_{i}(t,\mathbf{x} )\,, \\
	J_{ij} &= \int d^p  x\, \lp x_i p_j(t,\mathbf{x} ) - x_j p_i (t,\mathbf{x} )\rp\,,
\end{align}
are conserved under time evolution. These conservation laws not only provide insight on the dynamics, but also constitute a non-trivial check of our numerical procedure. 
In section \ref{sec:phys_quant} we will explain how they are related to the corresponding, appropriately normalized, physical quantities of a black hole. 

The conservation of the mass and momenta of the black brane, even under highly dynamical evolution, is possible due to the absence of gravitational wave emission when $D\to\infty$, a feature that we will analyze in detail in section \ref{sec:grav_rad}. Actually, the radiation is absent at all orders in $1/D$ since the suppression factor $\sim D^{-D}$ is non-perturbative \cite{Emparan:2013moa}.

Interestingly, the total entropy of the system is also conserved, since in the effective theory the horizon area density $a(t,\mathbf{x})$ is at all times and everywhere equal to the mass density $m(t,\mathbf{x})$. The proportionality constant between the entropy and mass densities is the temperature, which therefore remains fixed throughout time and space. However, this conservation of the entropy only holds at leading order at large $D$.\footnote{And then for neutral black branes, but not charged ones \cite{Emparan:2016sjk}.} The first corrections in $1/D$  generate entropy through squared shear terms $\sim \nabla_{(i} v_{j)} \nabla^{(i} v^{j)}$.\footnote{This was already clear from the fluid/gravity correspondence for asymptotically flat black branes \cite{Camps:2010br,Caldarelli:2012hy}, taken in the limit $D\to\infty$: the coefficient of the entropy-generating term in the fluid is $\propto 1/D$. The result has been explicitly verified within the large-$D$ effective theory in \cite{Herzog:2016hob,Dandekar:2016fvw,Bhattacharyya:2016nhn}.}

\subsection{Black holes, black bars, and their linear stability}

Henceforth we restrict ourselves to configurations in two spatial dimensions, $p=2$, with line element
\begin{equation}
	ds^2  = dr^2 + r^2 d \phi^2
\end{equation}
and momentum
\begin{equation}
	p = p_r dr + p_\phi d \phi\,.
\end{equation}

\subsubsection{MP black holes}

Eqs.~\eqref{eq brane1} and \eqref{eq brane2} can be solved to find singly-spinning MP black holes \cite{Myers:1986un} as Gaussian blobs on the brane, of the form
\begin{equation}
\label{BH soln}
	m(r) = m_0 \exp \left( - \frac{r^2}{2(1+a^2)} \right) , \qquad 	p_r = \partial_r m  , \qquad p_\phi = m \Omega r^2\,,
\end{equation}
where $m_0$ is an arbitrary constant and the rotation parameter $a$ determines the angular velocity, $v^\phi=\Omega$, as
\begin{equation}
	\Omega = \frac{a}{1+a^2}\,.
\end{equation}
The angular momentum per unit mass is
\begin{equation}
	J/M = 2 a\,.
\end{equation}

We can boost these blobs along the brane with constant velocity $u_i$ to set them in rectilinear motion, so that
\begin{align}
\nonumber
	m &= m_0 \exp \left( - \frac{(x_i - u_i t)(x^i - u^i t)}{2(1+a^2)} \right)\,, \\
\label{boostedBH}
	v_i &= u_i + \frac{a}{1+a^2} \varepsilon_{ij} (x^j - u^j t)\,.
\end{align}

Ref.~\cite{Andrade:2018nsz} showed that the Gaussian blobs \eqref{BH soln} not only reproduce the physical properties of stationary MP black holes, but they also exhibit their ultraspinning instabilities at large enough $J/M$ \cite{Emparan:2003sy,Dias:2009iu,Suzuki:2015iha}: these solutions of the effective theory are linearly unstable whenever $a \geq 1$.\footnote{The linear stability of MP black holes in the limit $D\to\infty$ was first studied in a different approach in \cite{Suzuki:2015iha}. The results were then reproduced in \cite{Andrade:2018nsz} using the black brane effective theory.} The onset of the instabilities is characterized by the appearance of marginal perturbation modes of the form
\begin{equation}
	\delta \Phi_A = e^{- i \omega t + i m_\phi \phi } \delta \hat \Phi_A(r)\,,
\end{equation}
\noindent where $\Phi_A = m, p_r, p_\phi$, with purely real frequencies that are co-rotating, \ie 
\begin{equation}
	\omega = | m_\phi | \Omega\,.
\end{equation}
These modes are present when
\beq
a^2= |m_\phi|+2k-1\,,
\eeq
where $k=0,1,2\dots$ is an `overtone' number corresponding to the number of `bumps' along the polar direction (\ie $r$) of the horizon.

The time-independent, axisymmetric perturbations with $m_\phi = 0$ appear when $k=2,3,\dots$, \ie when
\begin{equation}\label{aximodes}
	a= \sqrt{3}, \sqrt{5}, \sqrt{7}  \ldots \qquad\textrm{(axisymmetric marginal modes)} \,.
\end{equation}
Non-axisymmetric corotating perturbations with $|m_\phi| \geq 2$ occur for 
\begin{equation}\label{naximodes}
	a = 1, \sqrt{2}, \sqrt{3}, 2, \ldots\qquad \textrm{(non-axisymmetric marginal modes)} \,. 
\end{equation}
The latter are time-dependent perturbations, but they are marginal since their frequency is real. They are perturbations at the verge of an instability: increasing the rotation beyond the critical values above, one finds unstable modes that grow exponentially with time \cite{Andrade:2018nsz}. 

The non-axisymmetric marginal mode at $a = 1$ $(J/M = 2)$ is associated to a dipolar deformation with $|m_\phi| = 2$. It is of particular importance, since above this critical rotation the MP black holes are always unstable. Furthermore, this linear marginal mode marks the appearance of a branch of black bar solutions emerging from the black hole branch. They can be extended to a fully non-linear family of solutions, as we describe next.

\subsubsection{Black bars}\label{subsubsec:blackbars}

The effective equations have a simple non-axisymmetric, time-dependent, explicit solution, namely
\begin{equation}
\label{m bar}
	m = \exp \left[ 1 - \frac{r^2}{4} \left( 1 + \sqrt{1- 4 \Omega^2} \cos( 2(\phi - \Omega t) ) \right) \right]
\end{equation}
with constant $\Omega\leq 1/2$ and
\begin{equation}
\label{ps bar}
	p_r = \partial_r m , \qquad p_\phi = \partial_\phi m + \Omega r^2 m\,.
\end{equation}
For $\Omega=1/2$ this reduces to the MP black hole \eqref{BH soln} with $a=1$. Expanding \eqref{m bar} near $\Omega  = 1/2$ yields the marginal dipole perturbation of that black hole, proving the claim that \eqref{m bar} is the non-linear extension of that mode.

The spin per unit mass of these solutions is
\begin{equation}\label{baromega}
	\frac{J}{M} = \Omega^{-1}\,.
\end{equation}
Their interpretation as bars that rotate rigidly with angular velocity $\Omega$ becomes clearer using co-rotating Cartesian coordinates
\begin{align}
	x(t) &= x^1 \cos \Omega t + x^2 \sin \Omega t\,, \nn\\
	y(t)  &= x^2 \cos \Omega t - x^1 \sin \Omega t\,,\label{corot}
\end{align}
in which \eqref{m bar} becomes the oblong Gaussian profile
\begin{equation}\label{bar soln}
	m = \exp \left( 1 - \frac{x^2(t)}{2 \ell_\perp^2}  - \frac{y^2(t)}{2  \ell_{\|}^2}  \right)\,,
\end{equation}
with longitudinal and transverse axial lengths,
\begin{equation}
\label{ls bar}
	 \ell_{\|}^2 = \frac{2}{1- \sqrt{1- 4 \Omega^2}} \quad \geq \quad \ell_{\perp}^2 = \frac{2}{1+ \sqrt{1- 4 \Omega^2}}\,.
\end{equation}

Observe that, for a given mass, longer bars with higher spin rotate more slowly---a fact that will be very important when we discuss their emission of radiation. 
The bar becomes infinitely long when $\Omega\to 0$, with 
\beq\label{longbar}
\ell_{\|}\to 1/\Omega\,,\qquad \ell_\perp\to 1\,,
\eeq
and it approaches the static configuration of a black string. Since black strings are known to be unstable, with marginal modes at the threshold of the instability \cite{Gregory:1993vy}, it is natural to expect that these rotating bars are also unstable and display marginal co-rotating modes.

These modes were indeed found in \cite{Andrade:2018nsz}. 
They deform the bar along the longitudinal $y$ direction, and occur for 
\begin{equation}
\label{bar MM Omega}
	\Omega = \frac{\sqrt{n_y-1}}{n_y},  \qquad n_y = 3,4,5, \ldots \qquad\textrm{(black bar marginal modes)}\,. 
\end{equation}
The modes with even $n_y=4,6,8,\dots$ are symmetric deformations around the center of the bar. In this article we will focus mostly on these modes, the first of which appears at
\beq
\Omega=\frac{\sqrt{3}}{4}\simeq 0.43\,,\qquad \frac{J}{M}\simeq 2.31\,.
\eeq
This mode can be added or subtracted to the bar. The first possibility gives rise to a $(+)$-branch of deformed black bars pinched at the center, so they are dumbbell-shaped. Instead, in the $(-)$-branch the bar grows a bulge at its center and gets thinner at its ends, resembling now a spindle. This two-branch bifurcation (which is absent for black strings, since, in them, the two branches are merely translated versions of each other) is analogous to the one observed for bumpy black holes in \cite{Dias:2014cia,Emparan:2014pra}. The existence of the dumbbell $(+)$-branch is natural due to the centrifugal outward motion of mass, but the spindle $(-)$-branch is perhaps less intuitive. In the case of bumpy black holes, the horizons in this branch develop large curvatures at their outer edge, eventually diverging to a singularity in the rotation plane. It is reasonable to expect a similar fate of spindles.

The above discussion refers to stationary solutions, but in section~\ref{sec:BB_evol_inst} we will observe this double-branchedness in the time evolution of perturbed black bars: by reversing the sign of the initial perturbation, the bar increasingly becomes either dumbbell- or spindle-shaped. However, in the black bars that form in a black hole merger we never observe spindles; the collision configuration appears to naturally lead to the formation of dumbbells.

\bigskip

\begin{figure}[t]
\centering
\includegraphics[width=0.7 \linewidth]{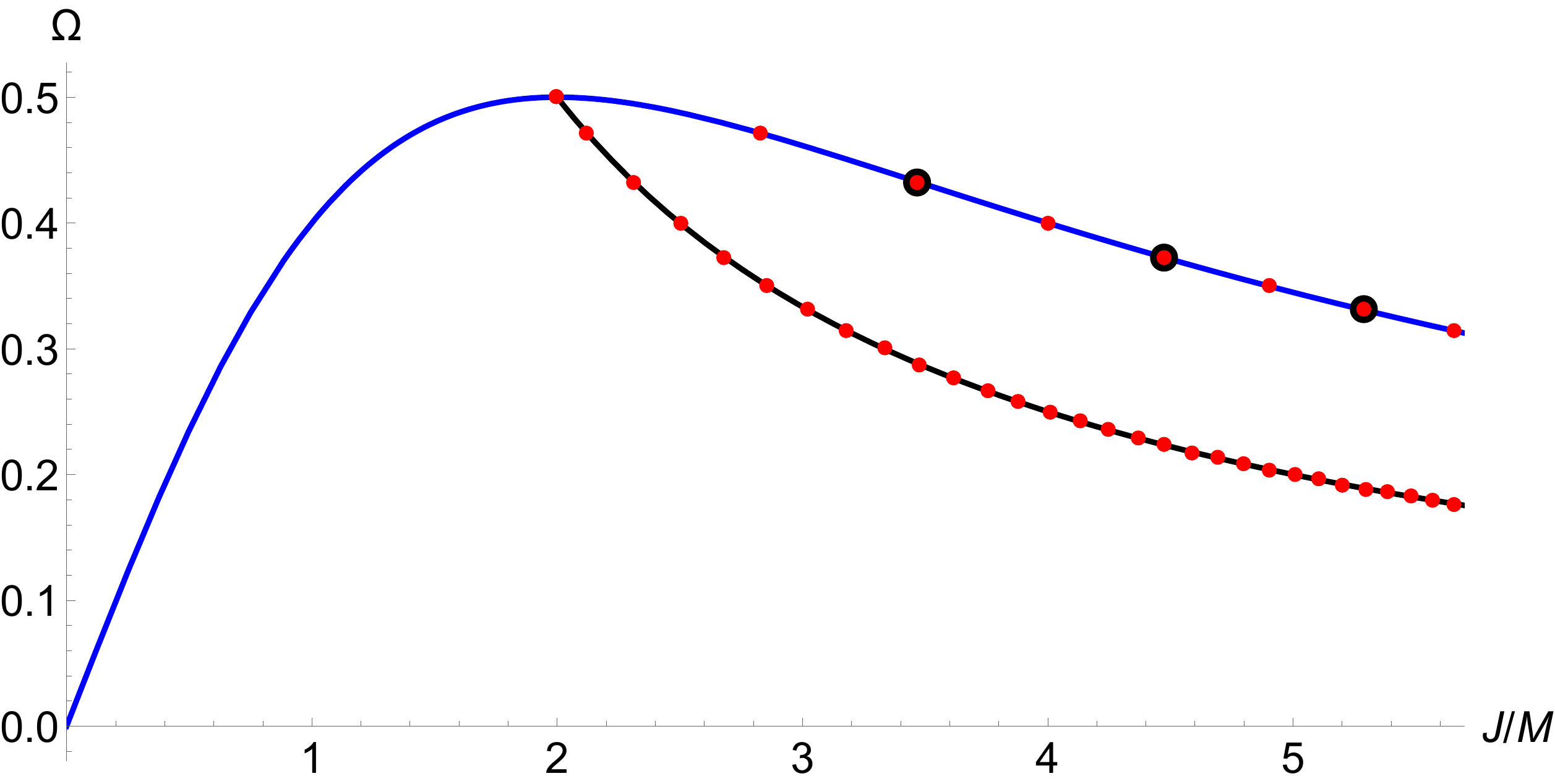}
\caption{\small Phase diagram of MP black holes (blue line) and black bars (black line). Red dots indicate the presence of non-axisymmetric marginal modes. Black circles indicate axisymmetric marginal modes of MP black holes. Observe that marginal modes of MP black holes and black bars appear at the same values of $\Omega$ but different $J/M$.  (Reproduced from \cite{Andrade:2018nsz})}%
\label{fig:pdiag}
\end{figure}

\begin{figure}[t]
\centering
\includegraphics[width=0.8 \linewidth]{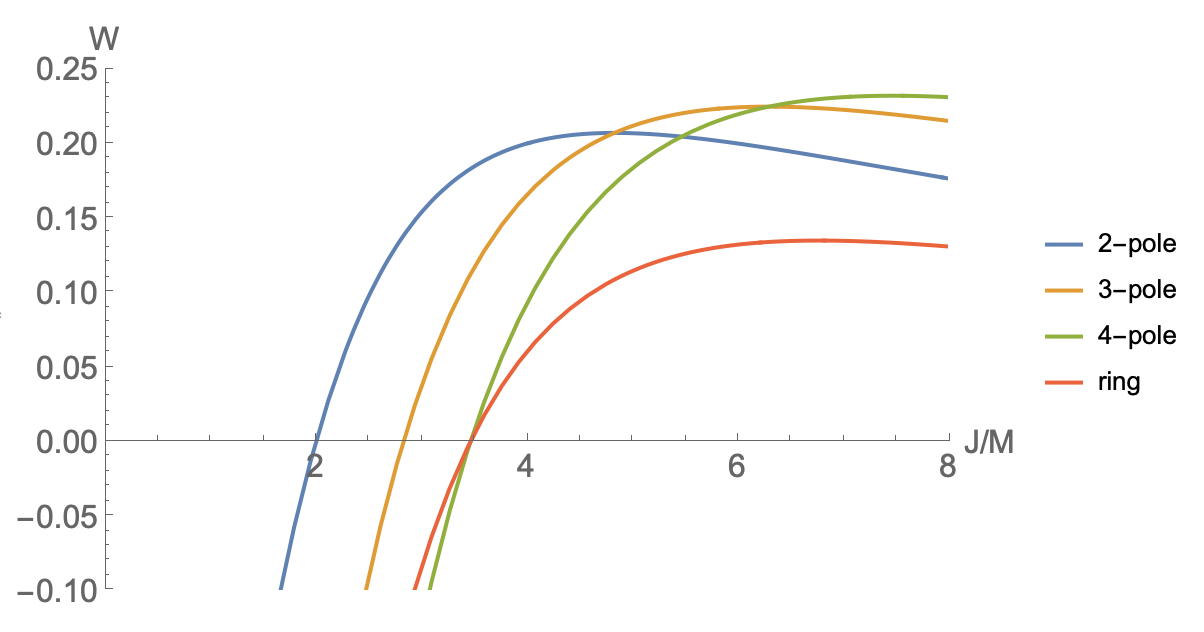}
\caption{\small Growth rates $W$ of unstable modes of MP black holes at $D\to\infty$, as a function of the spin per unit mass $J/M$ of the black hole (as calculated in \cite{Suzuki:2015iha,Andrade:2018nsz}). Higher non-axisymmetric modes (`$m_\phi$-poles') become successively dominant as the spin grows, and generically overwhelm the development of axisymmetric `ring' deformations.}%
\label{fig:instabrates}
\end{figure}

We summarize these results using two figures that will be very helpful for understanding our numerical simulations in the following sections. Figure \ref{fig:pdiag} is the phase diagram of stationary MP black holes and black bars: it will help interpret the final states of black hole collisions. Figure \ref{fig:instabrates} shows the growth rates of the unstable quasinormal modes of MP black holes (obtained in \cite{Suzuki:2015iha,Andrade:2018nsz}) as $J/M$ increases. It indicates which deformation is expected to dominate, at a given value of $J/M$, in the evolution of an unstable MP black hole subject to a generic perturbation.

\subsection{Physical magnitudes}
\label{sec:phys_quant}

A careful study of the matching of the solutions of the effective theory \eqref{largeD metric} to the geometry of an asymptotically flat black hole yields the physical mass, area, spin, angular velocity and surface gravity of the configurations (denoted in boldface), in terms of the effective theory magnitudes in section \ref{sec:eff_quant}, as
\beqa
\mathbf{M}&=&\frac{\Omega_{n+1}}{16\pi G} r_+^{n+2}\,\frac{n+3}{2\pi m_0} M\,,\\
\mathbf{A}&=&\Omega_{n+1} r_+^{n+3}\,\frac{1}{2\pi m_0} M\,,\\
\mathbf{J}&=&\frac{\Omega_{n+1}}{16\pi G} r_+^{n+3}\,\frac{1}{2\pi m_0} J\,,\\
\mathbf{\Omega}&=&\frac1{r_+}\Omega\,,\label{eq:Omegaphys}\\
\boldsymbol{\kappa}&=&\frac{n}{2r_+}+\frac1{2r_+}\ln m_0\,.\label{eq:kappaphys}
\eeqa
Here $n=D-5$, and $r_+$ is a length scale (necessary, since in the effective theory all quantities are dimensionless) that corresponds to the radius of the transverse sphere $S^{n+1}$, with unit volume $\Omega_{n+1}$, at the rotation axis. It can be eliminated in favor of $\boldsymbol{\kappa}$ in the expressions for the other physical quantities. A useful, equivalent form of \eqref{eq:kappaphys} is
\beq
m_0=\lp\frac{2\boldsymbol{\kappa}r_+}{n}\rp^n\,.
\eeq

Observe that we have distinguished $r_+^{n+2}$ in the mass from $r_+^{n+3}$ in the area and spin, even though they become the same as $n\to\infty$; we do this in order to maintain the correct dimensionality of these magnitudes. Other apparently subleading dependences at large $n$ have also been fixed through matching to known exact solutions---not a necessity but a convenience. The first correction to the surface gravity (and temperature)---which, as we mentioned earlier, is constant at leading large $n$ order---can be consistently determined from the geometry  \eqref{largeD metric}.

A convenient parameter to characterize the configurations is the spin per unit mass $J/M$. The corresponding dimensionless physical magnitude is
\beq\label{eq:JMphys}
\frac{J}{M}=\frac{D-2}{r_+}\frac{\mathbf{J}}{\mathbf{M}}\,,
\eeq
where the horizon radius $r_+$ was invariantly defined above.
This allows to translate in a simple manner the parameter $J/M$ of our colliding black hole simulations to the physical magnitudes in a collision at finite $D$.  Again, the factor $D-2$, instead of simply $D$, is chosen to better match known exact solutions, but it need not be accurate for generic configurations.

\section{Collision, merger, and break up}
\label{sec:bhcolls}

It is worth pausing to note what the effective equations \eqref{eq brane1}, \eqref{eq brane2} achieve for solving the fully non-linear, time-dependent evolution in a black hole merger and other similarly complex phenomena. The problem is straightforward: we can specify initial data corresponding to black holes (\ie blobs) moving towards each other with the aid of \eqref{boostedBH}, and simply follow the time evolution by numerically solving \eqref{eq brane1}, \eqref{eq brane2}. 
Since the equations are first order in time, we only need to supply the field configuration $(m, p_i)|_{t=0}$ in an initial time slice. There are no constraint equations to solve or keep track of, nor gauge issues: the analysis that led to the effective equations disposed of them already. We can directly read off the gauge-invariant, physical quantities of interest from the outcome of the time integration. Perhaps even more importantly, the effective dimensionality of the problem has been reduced by one, since the dependence on the radial variable $\rho$ away from the horizon  has been explicitly integrated in the effective theory. Then, our simulations in the $2+1$ dimensions of the effective theory correspond to $3+1$ evolutions in the complete spacetime (plus the $n+1=D-4$  dimensions of the ``passive'' $S^{n+1}$).

Interestingly, the effective equations are almost linear, with all the non-linearities confined to the last term in \eqref{eq brane2}. That is, this term alone is responsible for the interaction between the two colliding black holes: without it, they would pass through each other undisturbed. Both \eqref{eq brane1} and \eqref{eq brane2} resemble diffusion equations,\footnote{So the system evolves irreversibly even though the total entropy remains constant.} and  they produce very stable numerical evolution. Since there is no gravitational radiation at $D\to\infty$, no wave extraction is required (nor possible!) and the asymptotic behavior needs no special consideration. We simply impose periodic boundary conditions in a square domain in the spatial directions.

With all these simplifications, black hole collisions can be numerically simulated in not more than a few minutes in a conventional computer.

\subsection{Black brane as `regulator'}

An unavoidable feature of our effective theory of black holes is the presence of a black brane horizon at all points. The Gaussian profiles \eqref{BH soln} and \eqref{bar soln} extend all the way to infinity on the brane, and so the black holes are never completely localized---indeed, the presence of a non-vanishing mass and area density everywhere is a requisite for the validity of the effective theory.

Notice, though, that the mass density asymptotes to zero at infinity with exponential fall-off, so the infinite brane background is not a problem for the computation of extensive physical magnitudes of black holes (their total mass, area and spin), which never diverge. The Gaussian localization on the horizon is indeed a very basic feature of the large $D$ limit of spherical (or ellipsoidal) black holes \cite{Andrade:2018nsz}.
Observe also that, even if infinitely extended black branes of constant mass density are GL-unstable, this instability does not afflict the Gaussian blobs; actually, the blobs naturally appear as stable end states of the GL instability. We may say that the exponential fall-off leaves too little mass density at large distances to clump into smaller blobs.

Nevertheless, the ever- and omni-present horizon introduces peculiarities in configurations with more than one black hole. In an initial configuration with two blobs, the brane continuously connects them, and we cannot unambiguously say where one black hole ends and the other begins---at least not without introducing an arbitrary cutoff, \eg where the mass density becomes $10^{-2}$ of the peak density. And we cannot exactly determine either the moment when two black holes merge, since they are always part of one and the same continuous horizon.

More relevant to our purposes, the black brane horizon  in \eqref{largeD metric} is always regular and there never appears a singularity in it. Then, strictly speaking, within this approach we can never observe a violation of CC---an important point that we will return to in section \ref{sec:CCV}.  In our simulations, as we will see, two blobs approach and merge into a single one, which then splits into two different blobs that fly away from each other. But at all moments the horizon is smooth and continuous; we cannot ascribe a splitting instant without introducing an arbitrary cutoff at low mass densities on the brane.\footnote{A Gaussian blob can be matched to the pole region of a stationary black hole \cite{Andrade:2018nsz}, and the initial configuration in the effective black brane theory with two blobs would correspond to the transient, dumbbell-like black hole formed when two separate black holes touch and quickly merge in a timescale $\sim 1/D$. The results of \cite{Emparan:2019obu} (although in stationary configurations) suggest that the `neck' where this merger starts (or where the eventual break up occurs) affects only non-perturbatively in $1/D$ the physics in the pole region that is captured by the black brane effective theory. A more precise characterization of the separation between the two black holes would require going beyond the leading large-$D$ theory.\label{foot}}

Given the smoothness of all the evolutions in the effective theory, we may regard the `black brane background' as a kind of regulator in the system, which prevents the appearance of curvature singularities, and which allows the black holes to separate after a collision without the evolution ever breaking down. In this respect, the black brane may look similar to the apparent horizon regulator introduced in numerical holographic collisions in AdS \cite{Chesler:2008hg}. A distinction between the two is that in our effective theory the mass density asymptotes to zero, while \cite{Chesler:2008hg} introduce a small energy density everywhere. A more significant difference is that in \cite{Chesler:2008hg}, the regulator size can be parametrically separated from other scales, and therefore it is in principle possible to continuously remove it from the system, reducing its effects in a controllable manner. This is not possible in our set up.
Therefore, although the `brane regulator' does not impose any serious difficulty for the initial and intermediate stages in our black hole collisions, it does imply an inability to follow the evolution through to the putative horizon break up. This is a limitation inherent to the use of the black brane effective theory, and not merely a practical convenience for numerical solution, as it is in \cite{Chesler:2008hg} (and in our own handling of numerics, see appendix~\ref{app:method}).

The position we take is that our simulations do show that certain horizons (in collisions, and in the evolution of unstable black holes) develop instabilities that lead them towards localized pinch-offs---the evidence we present for this is clean and clear. For the further evolution of these horizons, we rely on what is known about the evolution of qualitatively related systems, in particular the development of black string instabilities. We will avail ourselves of all the current information about these in order to construct a strong, convincing case for the violation of CC. These conclusions should then be tested in future dedicated numerical simulations of collisions and black hole instabilities at finite $D$.

Now we can proceed to the results of our numerical simulations.

\subsection{Initial states}
\label{sec:in}

As stated above, in order to supply initial data for our simulations, 
we only need to specify an initial configuration $(m, p_i)|_{t=0}$. 
In the case at hand, we consider the superposition of two configurations 
of the form \eqref{boostedBH} centred at positions $\mathbf{ x}_1$, $\mathbf{ x}_2$, with initial 
velocities $\mathbf{ u}_1$, $\mathbf{u}_2$, mass parameters $m_{0,1}$, $m_{0,2}$ 
and rotational parameters $a_1$, $a_2$. 
Due to translational, boost and scale invariance of the effective equations, 
we can always set, say $\mathbf{ x}_1 = \mathbf{u}_1 = 0, m_{0,1} = 1$, without loss 
of generality, which shows that the space of parameters has dimension 7.

\begin{figure}[th]
\centering
\includegraphics[width=0.75 \linewidth]{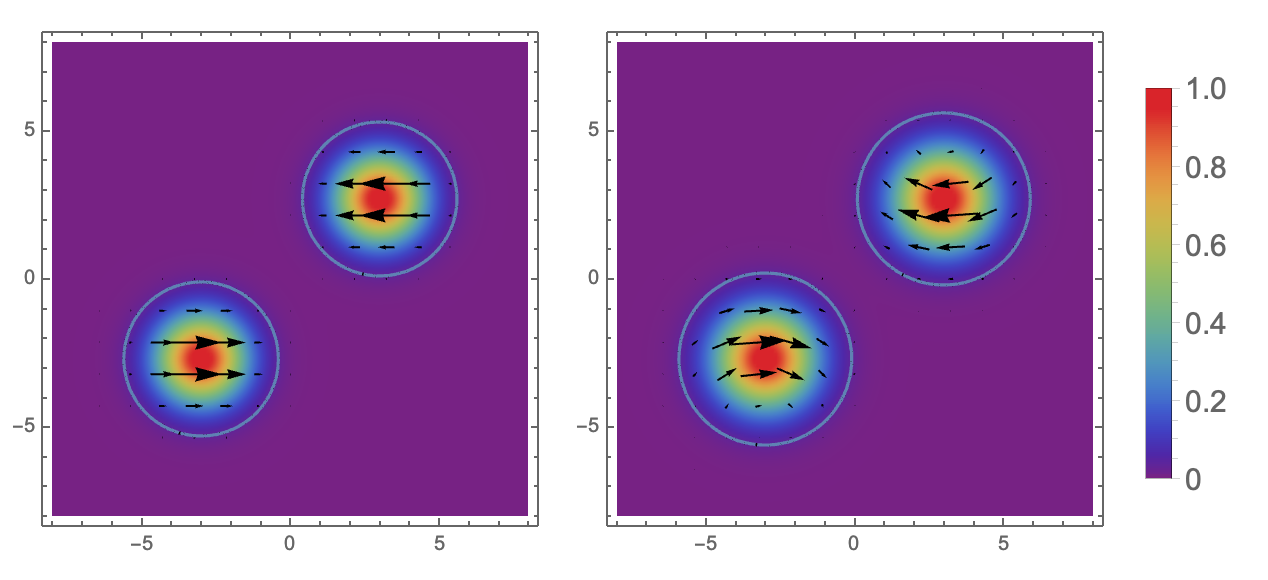}
\caption{\small Initial data for $u = 1$, $b = 2.5$, $x_0 = 3$, and $a = 0$ (left) and $a = 0.5$, $\sigma = 1$ (right).
Here we have chosen units $m_{0} = 1$.}%
\label{fig:IC}
\end{figure}

For numerical simplicity, we require our configurations to be reflection 
symmetric, which guarantees that the intermediate state will form and evolve 
at the center of our computational domain. We thus restrict ourselves to 
configurations of the form 
\begin{align}
	m_{0,1} & =m_{0,2} = m_0, \qquad a_1 = a = \sigma a_2 \\ 
	u_{x,1} &= -  u_{x,2} = u, \qquad  u_{y,1} =  u_{y,2}  =0,  \\
	x_1 &= - x_2 = - x_0  \qquad y_1 = -y_2 =  - b/2. 
\end{align}
Here, $\sigma = \pm 1$ controls whether the spins are aligned or 
anti-aligned. 
The parameter $x_0$ does not play any significant physical role, 
and it is chosen such that the Gaussian profiles do not overlap significantly 
at $t=0$. The initial states of our simulations are then characterized by three 
physical parameters: the relative velocity $u$, the impact parameter $b$ and the 
intrinsic spin controlled by $a$. 
In \cite{Andrade:2018yqu} we focused on configurations within this class with $a = 0$. 
We now expand our exploration of the space of parameters by considering
$a \geq 0$. 
We show examples of our initial conditions in figure \ref{fig:IC}.

\subsection{Black hole collision}

Having set up the initial conditions as described above, we follow the 
evolution of the system by numerically solving \eqref{eq brane1}, \eqref{eq brane2}.
We provide details of our numerical procedure in appendix \ref{app:method}, 
and in the remainder of this section we focus on the physical consequences. 

\subsubsection{Collision tomography}
\label{sec:tomo}

Before describing our results, we discuss a convenient way to characterize 
our final and intermediate state configurations. 
In this discussion, we will make heavy use of the symmetry assumptions 
described in section \ref{sec:in}, so our methods are only valid for simulations 
resulting from initial conditions within this class. 

A necessary condition for a configuration to be at equilibrium 
is that the maxima of $m(t, \mathbf{x})$ and $p_i(t, \mathbf{x})$ are 
constant in time. We use this simple criterion as a first check of the 
settling-down of a given time-evolving configuration. 
Moreover, in order to check whether a configuration 
evolves as a connected or disconnected object, we find it convenient to 
check for the value of energy density at the origin $m_{origin} := m(t, \mathbf{0})$. In all our 
initial states $m_{origin}$ is very close to zero, and this is clearly also the case 
for two black holes flying apart from each other.

In order to gain extra information about the spatial distribution of the 
energy density, we introduce the {\it tensor of inertia}\footnote{This differs from the conventional definition by a trace term, but it is equally good for our purposes.}
\begin{equation}
\label{inertia_t}
 I_{ij}(t) = \int d^2 x \, m(t, \mathbf{ x})\, x^i x^j\,.
 \end{equation} 
We can obtain the principal axes of rotation and effective lengths as eigenvectors
and eigenvalues $\lambda_1 \leq \lambda_2$ of the inertia tensor. Going to this frame of reference 
allows us to easily identify final states corresponding to MP black holes 
or bars by comparing the one-dimensional profiles along the axes with Gaussians.

\subsubsection{Final states}

Let us now discuss the final states obtained in our collisions. First, note that when
$b$ is sufficiently large, the black holes almost do not interact. This appropriately reflects the fact that at large $D$ the interaction between two massive objects decreases very quickly with the distance.\footnote{This is a little too glib. The range of the gravitational interaction between two separate black holes is $\sim 1/D$ (in units of the horizon radius), while the distances along the brane in the effective theory are much larger, $\sim 1/\sqrt{D}$. In the effective theory the direct gravitational attraction does not play any role, instead it captures well the elastic-type interaction between two black holes that have merged into a single horizon---see footnote \ref{foot}.}
All the interesting cases thus lie in a region in parameter space where $0<b < b_{crit}$, 
with $b_{crit}$ being a complicated function of $a$ and $u$. While it may be interesting to determine the shape of this function, we do not attempt it here.
We will assume below that we are always in the regime in which a non-trivial intermediate state 
forms, \ie in which a non-trivial interaction of the black holes takes place. 

A very important finding of \cite{Andrade:2018yqu} is that, at least for $a= 0$ in our highly symmetric
configurations, the type of final state that we arrive at does not depend separately 
on $u$ and $b$, but \emph{only on $J/M$, the total angular momentum per unit mass of the system}. 

Although we have found that this result holds for a large fraction of the parameter space, an extensive exploration reveals that,
for sufficiently high initial speeds, there are more exotic intermediate 
configurations in which additional lumps of energy emerge along the 
line that connects the two drifting black holes. This indicates that the scattering of black holes may not be only of the types
\beqa
2&\to& 1\,,\nn\\
2&\to& 2\,,\nn
\eeqa
that were considered in \cite{Andrade:2018yqu}, but more generally
\beq
2\to N=1,2,3,4\dots
\eeq
Such scatterings may be of interest, but their casuistics seems fairly complex. Henceforth we will assume that our initial speeds are small enough that the scattering is always $2\to 1$ or $2\to 2$.

For relatively small initial speeds---with our without initial intrinsic spins---, 
the available final states for a given value of $J/M$ can be predicted simply 
from the stability properties of the black holes and bars in the phase 
diagram in figure  \ref{fig:pdiag}.
More concretely, for $J/M < 2$ the final states are MP black holes, 
while for $2 < J/M < (J/M)_{crit} = 4/\sqrt{3} $ the collisions form 
bars. This critical value of $J/M$ corresponds to the first parity-even 
marginal mode of the bars in \eqref{bar MM Omega}: the dumbbell deformation of the bar. 
For $J/M$ above $(J/M)_{crit}$, the intermediate state formed in the 
collision breaks up. As we elaborate later, this is the signal of evolution towards a violation of CC. 

We have checked that introducing spin $a>0$ opens a new channel for 
violation of CC. Most notably,  we will show that the intermediate 
state of collisions of spinning black holes can be quantitatively accurately 
approximated as almost stationary black bars. This 
confirms that the mechanism responsible for their break up is the GL instability.

We have not made a detailed analysis of what is the precise final state after the horizon break up, \ie what is the outgoing impact parameter and the spins and velocities of the outgoing black holes, and how strongly these depend on initial state properties other than the total $J/M$. But it is possible to extract some generic features. In particular, we expect that the spin of the final black holes is smaller than that of the initial black holes, since when the horizon breaks up, it does so from a fairly long black bar, and therefore the outgoing impact parameter is larger than the initial one. Then, the final orbital angular momentum is larger than the initial one, and so the intrinsic spin must be smaller. For instance, this is visible in figure \ref{fig:wholething}, where the initial black holes look larger than the final ones, even though they necessarily have the same mass, but the size of the blobs, for a given mass, is larger for larger spin. With larger initial impact parameter or velocities the situation is less clear, and probably the initial and final spins are more comparable.

\begin{figure}
\centering
\includegraphics[width=0.49 \linewidth]{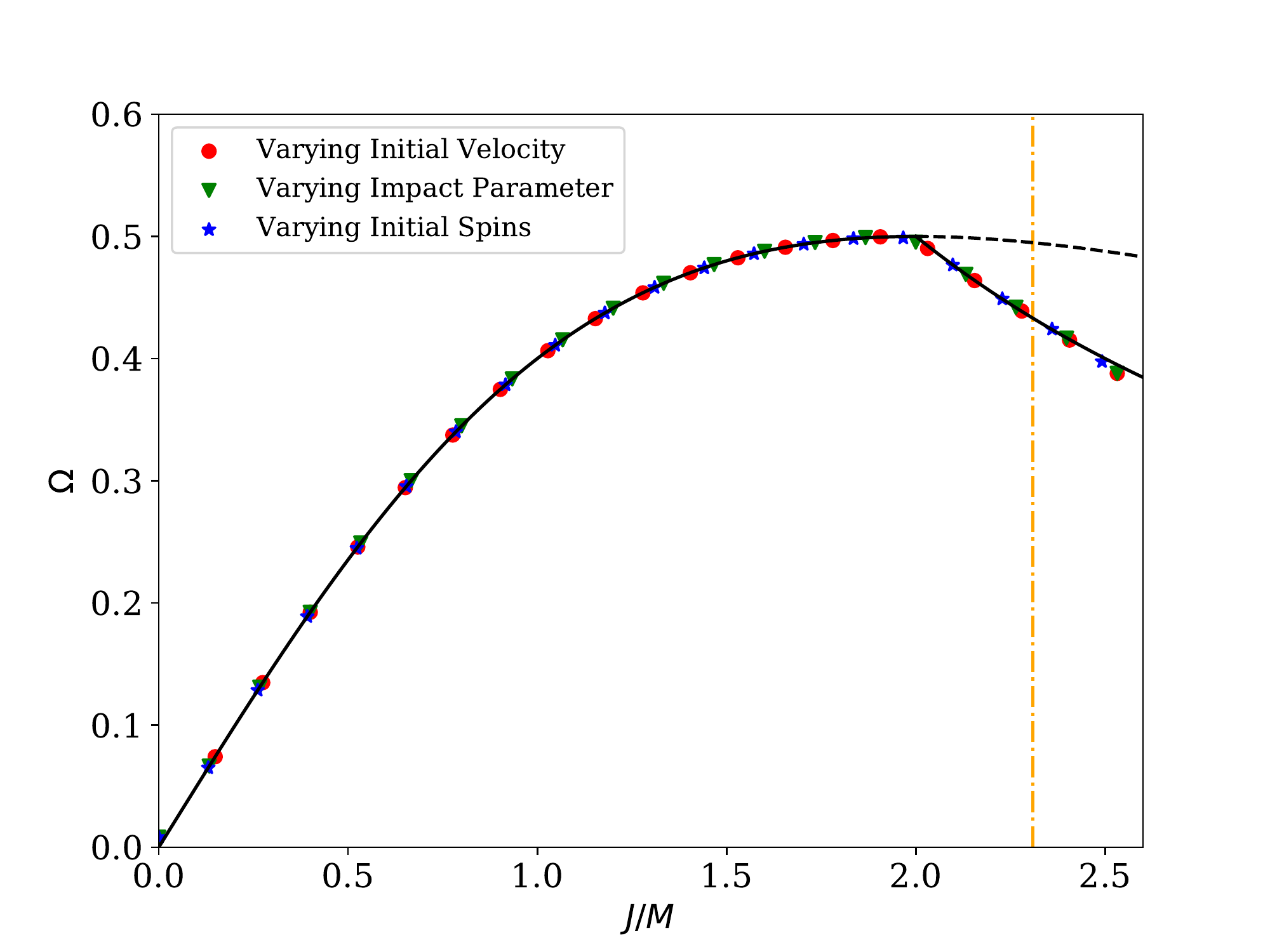} \includegraphics[width=0.49 \linewidth]{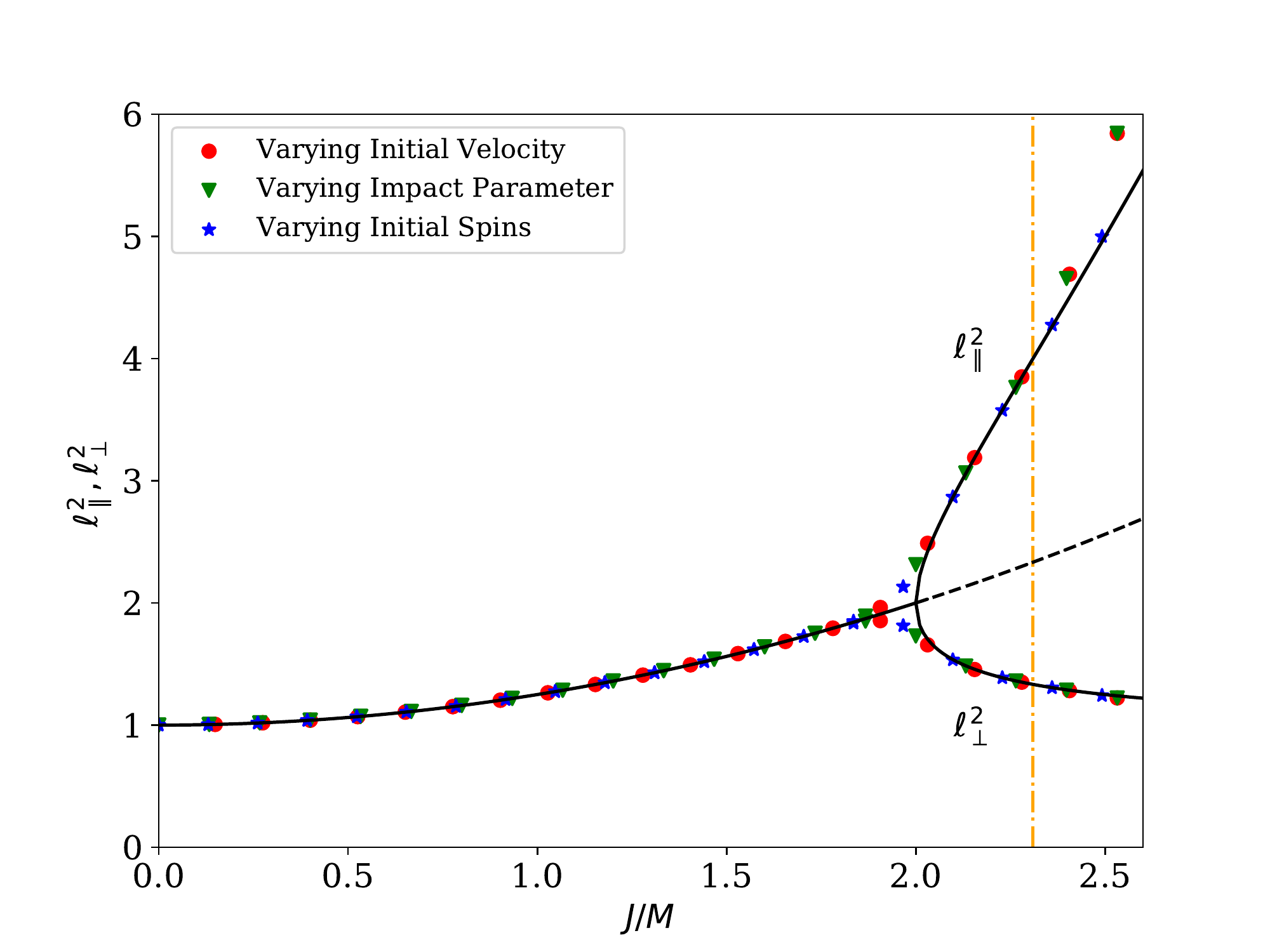}

\caption{\small Final states of collisions for varying impact parameter, initial velocities and spins. The dashed and continuous lines correspond to stationary MP black holes and black bars. The left plot  is superimposed on the phase diagram of  figure \ref{fig:pdiag}. The right plot shows the (squared) longitudinal and transverse axial lengths $\ell^2_\parallel$ and $\ell^2_\perp$ described in (\ref{bar soln}) of the final states, and provides more detailed evidence that the collision forms a rotating black bar. The values of $\ell^2_\parallel$ and $\ell^2_\perp$ are obtained by linear regression of $\log m$ with respect to $r^2$ along the longitudinal and transverse directions, respectively. We see that the initial value of $J/M$ (conserved along the evolution) predicts the end state. The vertical dash-dotted orange line shows the onset of the fundamental symmetric mode of the black bar. We observe certain configurations
lying beyond this line, but these are not stable: their appearance is an artifact of running the simulation for a finite amount of time.}%
\label{fig:final_states_omega}
\end{figure}

In Figure \ref{fig:final_states_omega} we show the final states of simulations with parameters in the following range:

\begin{itemize}
\item Varying Initial Velocity: $u \in [0.1, 1.7]$, $b = 3$, $x_0 = 3$, $a = 0$
\item Varying Impact Parameter: $u = 1.7$, $b \in [0, 3]$, $x_0 = 3$, $a = 0$
\item Varying Spins: $u = 0.5$, $b = 0$, $x_0 = 5$, $a \in [0, 1.25], \sigma = 1$
\end{itemize}

Within this domain in parameter space, whether the collision is $2\to 1$ or $2\to 2$ can be predicted from the sole knowledge of the initial value of the total $J/M$ (which is conserved along the evolution).

\subsubsection{Intermediate black bars}
\label{sec:interm}

As discussed above, the intermediate states of the collisions that yield CC violation 
have the form of bar-like objects. This, combined with the fact that no stable bars 
are observed to form above the threshold predicted by the marginal mode computation, 
suggests that the mechanism for the pinching off is the GL instability present in the black bars.
However, there is a possible caveat: in the $a = 0$ simulations the intermediate
states are highly distorted elongated objects, which makes the comparison with 
actual bars rather indirect. 
It turns out that this argument can be put on a quantitative basis in the case of initial 
spinning black holes, as we now show in detail. 

\begin{figure}[th]
\centering
\includegraphics[width=0.5 \linewidth]{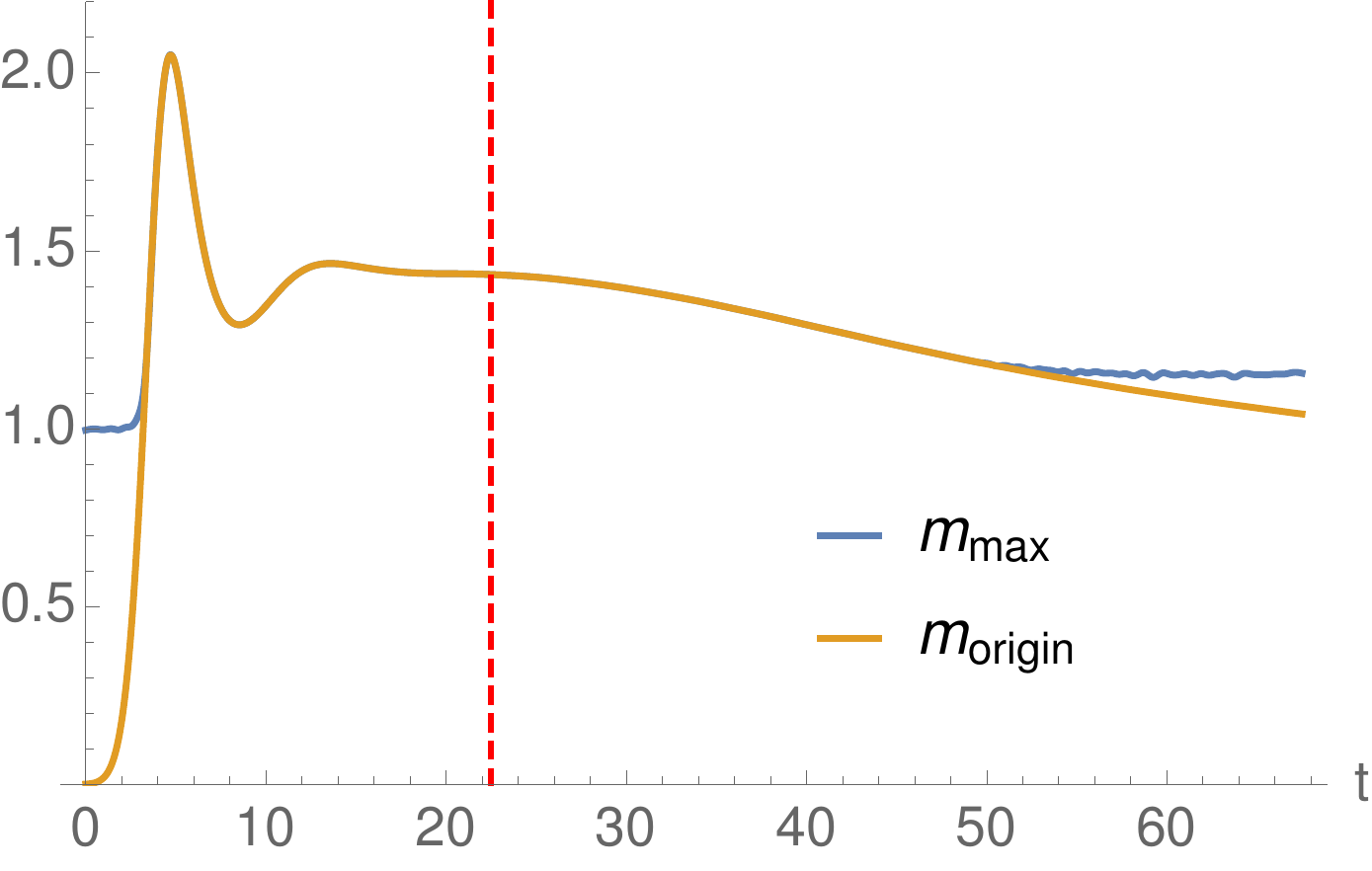}
\caption{\small Time evolution of the maximum of the energy density $m_{max}$ and 
its value at the origin $m_{(0,0)}$. After a short period, these two values become equal, 
signaling rigid rotation. At late times, their values begin to differ, as a consequence of 
the break up of the bar. We extract the profiles at a time $t \approx 22$, shown in red, 
and plot them in Fig. \ref{fig:bar_profiles} below.}%
\label{fig:mvst}
\end{figure}

Let us consider for concreteness the case of a collision with $a = 0.99$, $b=1.2$, $u = 1.2$. 
The energy density of the black holes is such that $m_0 = 1$. For these parameters, $J/M = 2.69$,
so we expect the formation of a bar-like object which should then break apart since we are in the 
region of unstable bars. This is indeed the case, as we show in Fig. \ref{fig:mvst}, where we depict 
the time evolution of the maximum of the mass density and its value at the origin. 
We see that $m_{max} = m_{(0,0)}$ for a long part of the evolution, which corresponds to the 
interval in which a metastable bar exists. At later times,  $m_{max} > m_{(0,0)}$, indicating
that the break up has taken place. 

\begin{figure}[th]
\centering
\includegraphics[width=1.0 \linewidth]{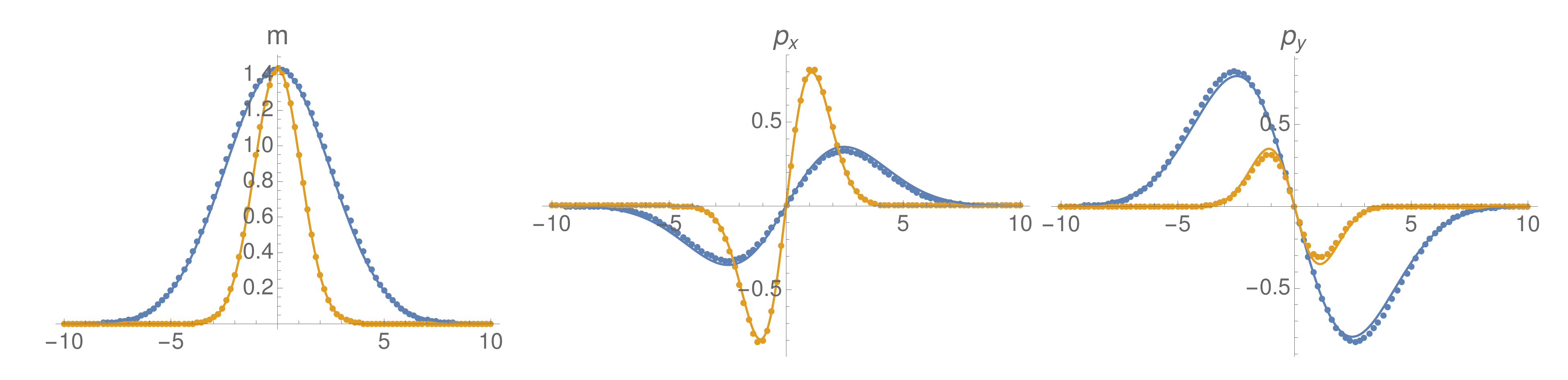}
\caption{\small Values of the profiles at $t \approx 22$ along the principal axes, in blue and yellow. 
The data points correspond to the numerical data, while the solid lines show the analityc profiles in 
\eqref{m bar}, \eqref{ps bar}. The value of $\Omega$ is obtained from our initial data, while $m_{max}$ 
is extracted from the numerics.}%
\label{fig:bar_profiles}
\end{figure}

We show the profiles of the energy and momentum at a time in which there is rigid rotation
in  Fig. \ref{fig:bar_profiles}. To ease visualization, we only show these at the principal axes, 
as defined by the inertia tensor \eqref{inertia_t}. We compare these to the values of an analytic
bar, with $\Omega$ given by the initial data $\Omega = M/J \approx 0.37$ and $m_{max}$
extracted from the numerics. 
We observe that there is excellent agreement
between the configuration and that of an analytic black bar. 

\subsection{Violation of Cosmic Censorship and subsequent evolution}\label{sec:CCV}

\subsubsection{Singularity formation}

We now have proven that the collision of two black holes at very large $D$, and with high enough total angular momentum, forms an intermediate, elongated, bar-like horizon that then pinches at its middle. However, as we discussed above, in the effective theory at $D\to\infty$ the pinch never shrinks to zero size, since at any finite distance on the brane there is always a non-zero thickness of the horizon. If $1/D$ corrections to the effective theory were included, their effects would grow large as the pinch becomes thinner, rendering the large-$D$ expansion inappropriate as an approximation to finite values of $D$. In other words, the large-$D$ approach employed in this article does not by itself allow to reveal the formation of a naked curvature singularity.

Nevertheless, as argued in \cite{Andrade:2018nsz} and further elaborated in this article, the instability of the intermediate bar state is, at least at its onset, of the same kind as the GL instability of black strings, even quantitatively (see section \ref{sec:bardecay}) . We can then draw upon the numerical simulations at finite $D$ of the non-linear evolution of the GL instability of black strings in \cite{Lehner:2010pn} (and of related higher-dimensional black holes in \cite{Figueras:2017zwa,Figueras:2015hkb}), which convincingly showsthe formation of a naked singularity at a horizon pinch in a finite time. The only effect that we envisage as possibly preventing a similar evolution of the black bar is a spin-down back to stability through gravitational radiation emission. This will be the subject of detailed study in section~\ref{sec:grav_rad}, where we show that this emission is very strongly suppressed for long bars, and also as $D$ grows. Thus, the conclusion seems to us inescapable that at large enough $D$ the merger with an intermediate long bar will end up producing a naked singularity.

The point may be raised that the numerical simulations in \cite{Lehner:2010pn} were performed in $D=5$ (and other relatively low $D$ in \cite{Figueras:2017zwa,Figueras:2015hkb}), while it is known that above a critical dimension $D_*\simeq 13.6$, an unstable black string may evolve into a stable non-uniform black string, instead of proceeding to a singular pinch \cite{Sorkin:2004qq,Emparan:2018bmi}. This, however, is not relevant to our analysis, since it is a consequence of the confining effect of the compact circle that the black string lives in. In any finite $D$, if the circle is long enough---compared to the thickness of the black string---the non-uniform strings are unstable and the evolution will not stop at them but proceed to pinch off \cite{Emparan:2018bmi}. In the case of a rotating black bar in asymptotically flat space, there is no limit to the distance to which it can spread, and thus there is nothing to stop its unstable evolution towards pinch-off. Furthermore, we expect that the centrifugal repulsion will accelerate the pinching faster than in the case of a black string. Indeed, it may well proceed quickly enough as to prevent the formation of the small `droplets' that were observed in \cite{Lehner:2010pn}.\footnote{A small central droplet at the rotation axis may appear, as we have observed in collisions with high enough velocities.}

So we see no plausible alternative to the conclusion that, at high enough $D$, if the unstable bar forms, a pinch in the horizon will develop where the curvature grows arbitrarily large in a finite time: a violation of CC.

\subsubsection{Proposal for resolution: Neck evaporation}

In our simulations, the `brane regulator' allows us to follow the evolution of the pinching bar and observe two blobs flying apart. However, as we argued, this regulator cannot be removed in a parametrically controlled manner from the theory. As a consequence, this part of our simulations of the system cannot be regarded as `proof' of how the process unfolds after the singular pinch-off. When General Relativity breaks down, the further evolution requires new laws of physics, arguably a quantum theory of gravity.\footnote{It may well happen that, if string theory is valid and the string coupling constant is small, General Relativity is replaced by classical string theory that resolves the singularity before reaching the Planck scale. The picture that we propose does not substantially change: the neck would evaporate at the Hagedorn scale instead of the Planck scale.} Here we want to propose a plausible resolution of the singularity such that the input from quantum gravity is minimal and affects very little the subsequent evolution of the system.

The neck that forms in the horizon has very high curvature, and may be regarded as a small, `Planck-size black hole', with very high effective temperature. It seems natural to expect that such an object, without any conserved charges that could prevent its decay, must indeed quantum-mechanically decay by emitting a few Planck-energy quanta, in a few Planck times. That is, we propose that the neck evaporates in much the same manner as the neck that forms in a fluid-jet evaporates (literally) and breaks the jet into a number of droplets. This break-up is not described by classical hydrodynamics, but rather by molecular dynamics; however, hydrodynamics quickly resumes control of droplet evolution after the brief episode of evaporation. Similarly, classical General Relativity resumes after the horizon breaks up, and controls how the two resulting black holes fly apart.

Note also that, although the evaporation of the neck is reminiscent of the expected endpoint of Hawking evaporation, in the case of a black string the evolution towards the Planck-size object is governed by classical dynamics and therefore is unaffected by the unitarity paradox.

If this picture is correct, then the evolution of the black hole collision and merger will result in a horizon pinch, which then quickly evaporates through quantum-gravity effects (just a little `pixie dust') and yields two outgoing black holes. The loss of classical predictivity is very small: the horizon bifurcates with a variation of the horizon area (increase or decrease) of only Planckian-size, and the uncertainty in the outgoing scattering angle will be proportional to at most a power of  $(M_\textrm{Planck}/M)$, where $M$ is the total mass of the system. Hence, the indeterminacy is a parametrically very small number for any macroscopic initial mass. Predictivity of the entire evolution using General Relativity will be maintained to great accuracy. Except for the details of the break up, the picture we have presented in figure~\ref{fig:wholething} will then be essentially correct.

\section{Non-linear evolution of ultraspinning black hole instabilities}
\label{sec:MP_evol_inst}

Our setup also allows the efficient simulation of the non-linear evolution of instabilities, such as those of ultra-spinning MP black holes \cite{Emparan:2003sy}. We will not attempt a detailed quantitative study, but rather a preliminary qualitative investigation of which intermediate and end states appear in the evolutions.

The development of the instabilities is quite different depending on whether they are triggered by axisymmetric or non-axisymmetric perturbations. In generic cases, the latter will dominate the evolution of an unstable black hole (see figure~\ref{fig:instabrates}). We find that the unstable black hole sheds off its `excess' angular momentum (\ie the spin above the stability limit of MP black holes) by breaking off smaller black holes. In the terminology of \cite{Emparan:2003sy}, this is `death by fragmentation', since `death by radiation' is outlawed in $D\to\infty$. We expect (see section \ref{sec:grav_rad}) that at large enough $D$ and large enough spin this violent, CC-violating chastisement of overspeeding black holes also prevails over the milder, CC-preserving radiative correction to stable, lower spin states. As $D$ increases, the value of the spin for which the CC-violating evolution occurs becomes smaller.

As a first case of interest we study non-axisymmetric perturbations of an ultraspinning black hole with $a=3$ . For each multipole $m_\phi$ we consider the fundamental mode (with the least nodes in the polar direction $r$, \ie $m_\phi\neq 0$ and $k=0$ in \cite{Andrade:2018nsz}, which first appear as marginal modes at the values \eqref{naximodes}). The first one is the dipole $m_\phi=2$ mode, with similar dynamics as for the collision process, but in this example at a very large spin $J/M=6$. The dipolar perturbation leads to the formation of an elongated horizon that resembles closely a stationary black bar. It then quickly decays into two smaller black holes after a formation of a dip in its middle.
\begin{figure}[h]
	\centering
	\includegraphics[width=0.225\linewidth]{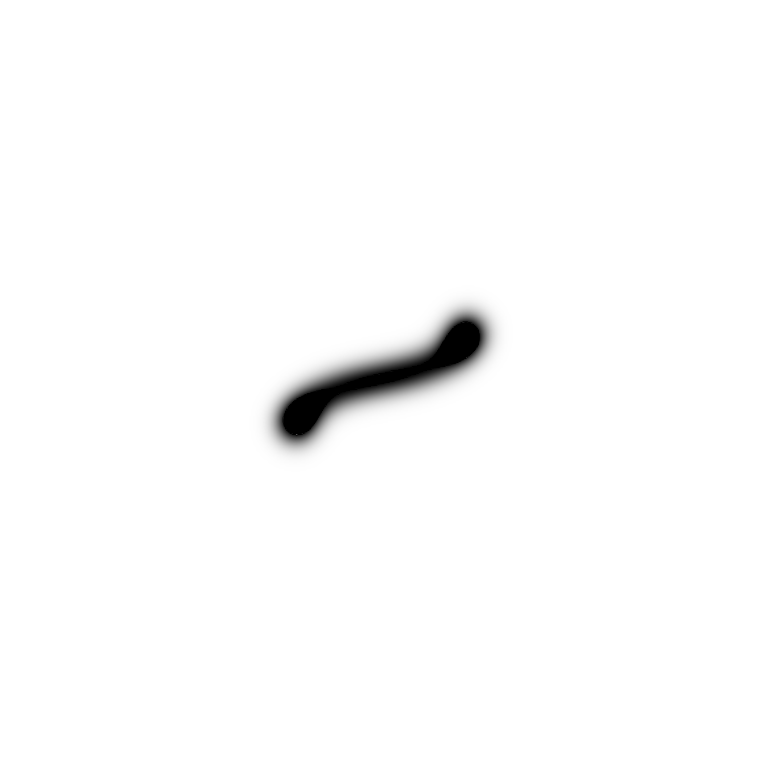}
	\includegraphics[width=0.225\linewidth]{fig/tripole}
	\includegraphics[width=0.225\linewidth]{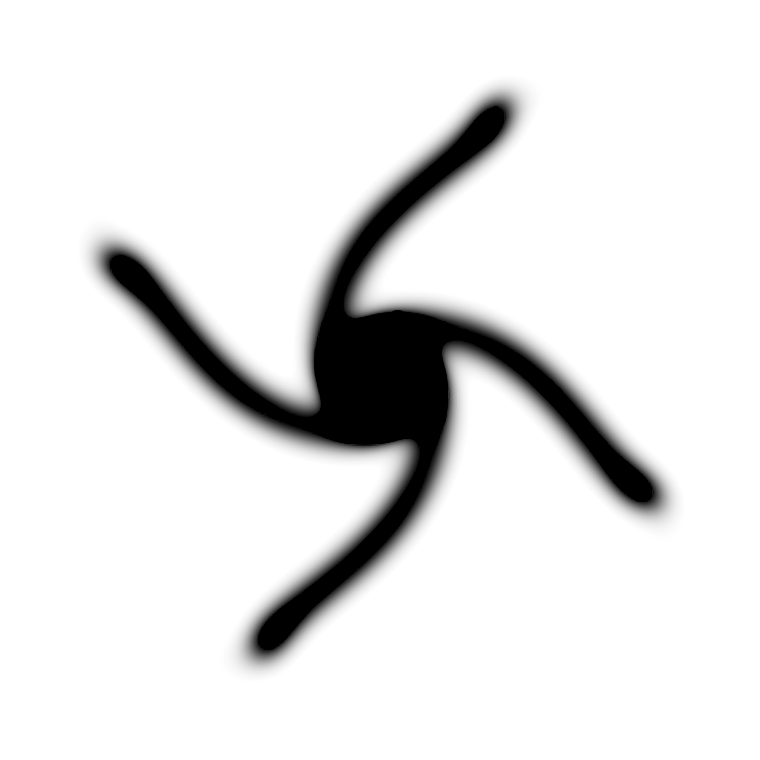}
	\includegraphics[width=0.225\linewidth]{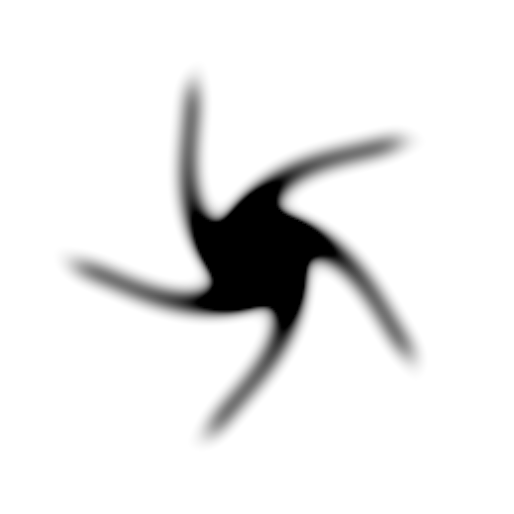}
	\caption{\small Intermediate states of the evolution of the MP black hole with $a=3$, after perturbations with $m_\phi=2,3,4,5$ (from left to right). These are again high-contrast density plots, where this time the minimal value that appears as black was chosen low enough to highlight the full structure, \ie these plots do not represent well that the relative mass density, which is higher in all places where blobs form.}%
	\label{fig:MPMultipoleDecay}
\end{figure}

For perturbations with $m_\phi \geq  3$, a novel set of intermediate states appear, which grow `arms' as shown in figure \ref{fig:MPMultipoleDecay}.  While growing longer and thinner during evolution, these arms develop a GL-like instability on their own and pinch off, leaving behind a number $m_\phi$ of small black holes that get slung away from the central MP black hole, which now has a spin within the stability bounds.

\begin{figure}[h]
	\centering
	\includegraphics[width=0.225\linewidth]{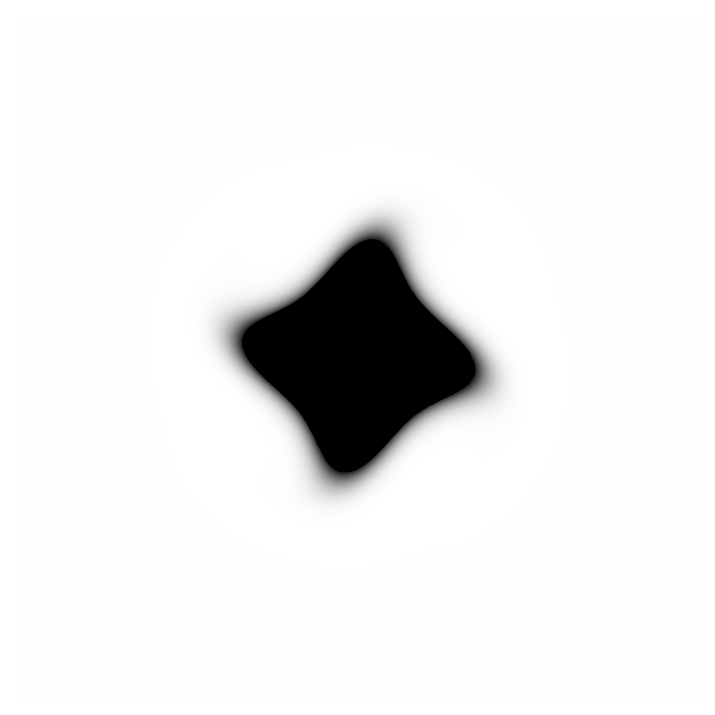}
	\includegraphics[width=0.225\linewidth]{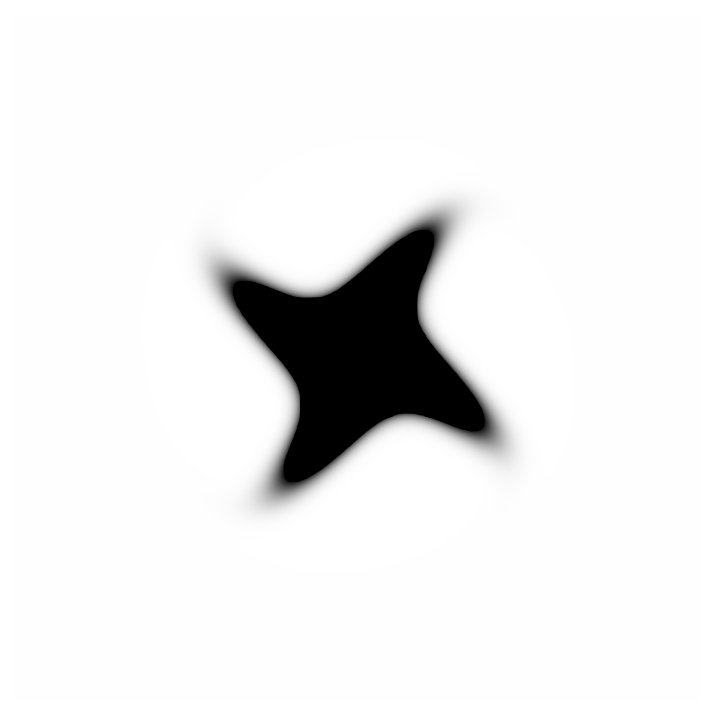}
	\includegraphics[width=0.225\linewidth]{fig/quadrupoleSnap3}
	\includegraphics[width=0.225\linewidth]{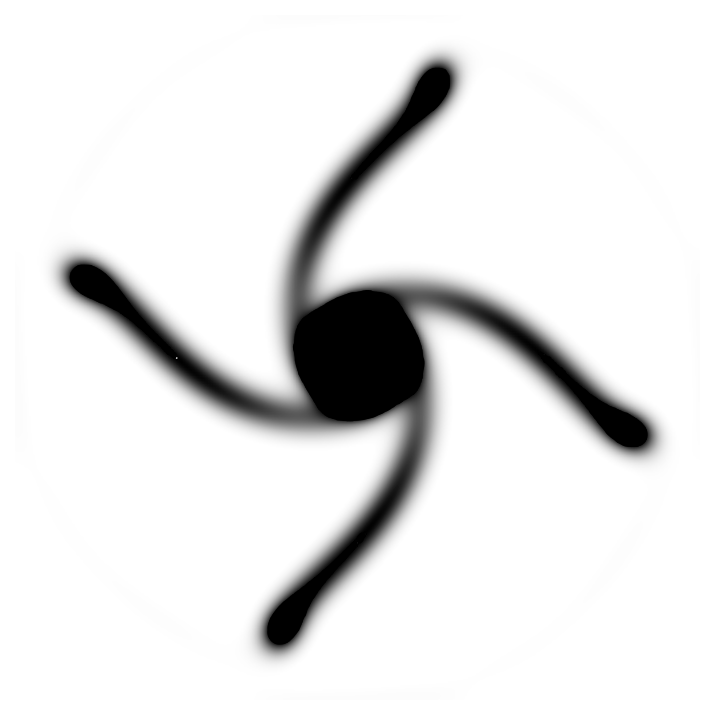}
	\caption{\small Snapshots of time evolution for the MP black hole with $a=3$, perturbed by the fundamental  $m_\phi=4$ mode}
	\label{fig:QuadrupoleEvolution}
\end{figure}

Figure \ref{fig:QuadrupoleEvolution} shows snapshots of the evolution in the case of a perturbation with $m_\phi=4$. These are strikingly similar to the images presented in \cite{Bantilan:2019bvf} for the evolution of MP black holes at spins high enough to excite the unstable quadrupole mode (see their figure 6).

\begin{figure}[h]
	\centering
	\includegraphics[width=\linewidth]{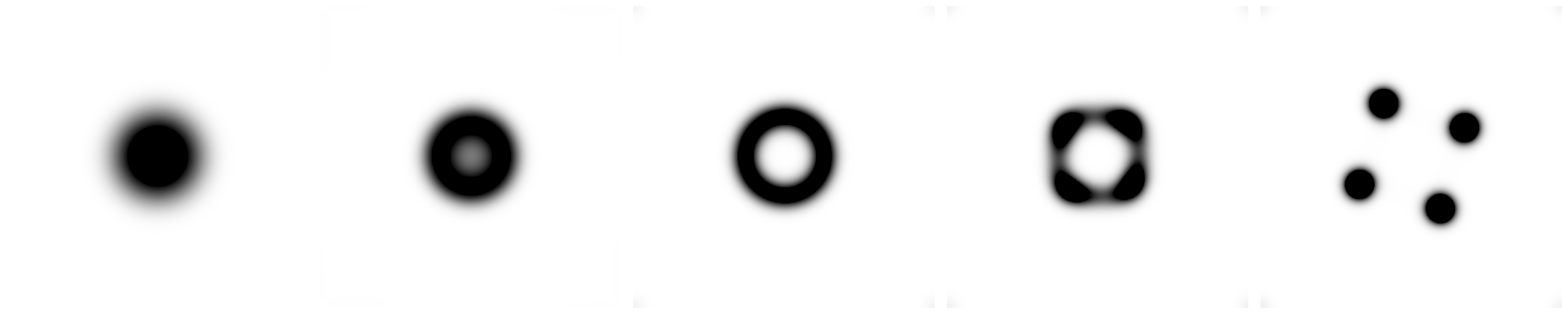}
	\caption{\small Snapshots of time evolution of a ring-like axisymmetric perturbation of the MP black hole with $a=3$. The eventual breakdown of axial symmetry is triggered by any generic perturbation, such as numerical noise.}%
	\label{fig:MPRingDecay}
\end{figure}

Next  we examine the first axisymmetric perturbation (which appears as a marginal mode at $a=\sqrt{3}$ \eqref{aximodes}, \ie $m_\phi=0$ and $k=2$ in \cite{Andrade:2018nsz}) for the same MP black hole with $a=3$ added with a positive amplitude. Figure \ref{fig:MPRingDecay} shows snapshots of the evolution. We find that the instability leads to the formation of a black ring. This was  observed in the axisymmetric numerical evolutions in $D=6$ of \cite{Figueras:2017zwa}. However, the axial symmetry of these rings is expected to break down by non-axisymmetric GL-like instabilities along the ring, and we do see this phenomenon: the black ring decays following a quadrupole perturbation that is triggered by numerical noise. Again, this agrees with the instability of thin black rings observed in \cite{Figueras:2015hkb}. Since our method allows us to evolve past the pinch-off we can observe four spherical black holes flying apart as the end state of the process.

Lastly, we consider the possibility of adding the above marginal mode, $m_\phi=0$ and $k=2$, again for a MP black hole with $a=3$, now added with an opposite (negative) amplitude, \ie a $(-)$-branch evolution. We observe, that instead of forming a dip in the middle of the MP black hole, one or more ringlike objects are emitted that move rapidly outwards and do not break up for the duration of our simulation.

\begin{figure}[h]
	\centering
	\includegraphics[width=0.25\linewidth]{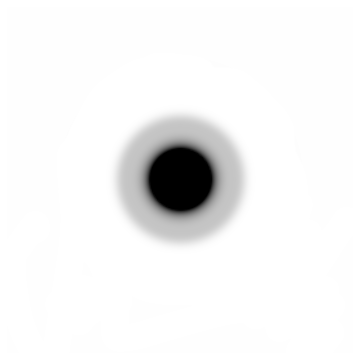}
	\includegraphics[width=0.25\linewidth]{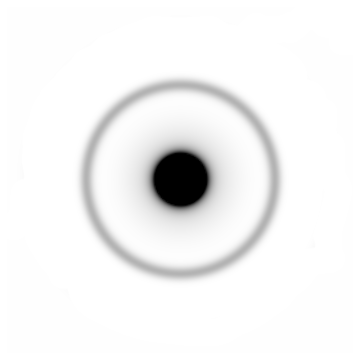}
	\includegraphics[width=0.25\linewidth]{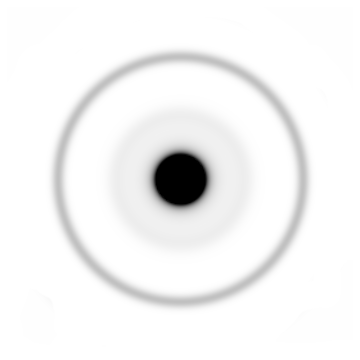}
	\caption{\small Snapshots of time evolution of a `negative' ring-like axisymmetric perturbation of the MP black hole (with rotation parameter $a=3$). The initial perturbation has amplitude opposite to the one in figure \ref{fig:MPRingDecay}, \ie a bulge at the center, instead of a pinch. }%
	\label{fig:MPposkDecay}
\end{figure}

\section{Black bar instabilities}
\label{sec:BB_evol_inst}
As for MP black holes, we can follow the non-linear evolution of the instabilities of black bars. 

\subsection{Spindles and Dumbbells}

\begin{figure}[th]
	\centering
	\includegraphics[width=0.30\linewidth]{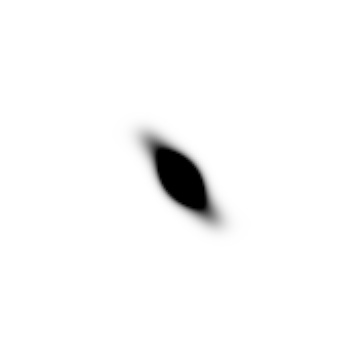}
	\includegraphics[width=0.30\linewidth]{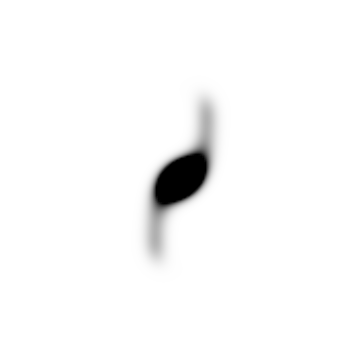}
	\includegraphics[width=0.30\linewidth]{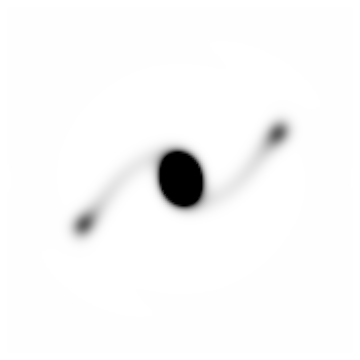}
	\caption{\small Snapshots of time evolution of the fundamental $n_y=4$ mode added with negative amplitude (i.e., creating a bulge in the middle, instead of a pinch). }%
	\label{fig:barspindleperturbation}
\end{figure}

As mentioned in section \ref{subsubsec:blackbars}, the fundamental symmetric mode of black bars  is of most relevance to us. It gives rise to two branches of solutions in phase space, depending on the sign of the amplitude with which the perturbation is added. Let us first consider the negative sign, which (by our conventions) leads to a spindle-like deformation. Figure \ref{fig:barspindleperturbation} shows snapshots of the evolution. These spindles resemble horizon shapes observed in \cite{Bantilan:2019bvf} in the numerical evolution of dipolar MP instabilities in $D=6,7$. The formation of pointy tips is followed by the development of arms, similar to the ones observed above for MP black holes. The arms subsequently pinch off, sending away two small black holes.

Since the qualitative evolution of the fundamental symmetric mode has already been described in earlier sections, here we will only estimate instability rates for this mode and the total duration of break-up.

\subsection{Black bar decay rates}\label{sec:bardecay}

We expect the unstable modes of the black bars to behave as 
\begin{equation}
	\delta \Phi_A = e^{W t} \delta \hat \Phi_A(r, \phi - \Omega t)\, ,
\end{equation}
where $\Phi_A = m, p_r, p_\phi$, and $W$ is the purely real instability rate. It is possible to estimate $W$ close to the onset of the zero modes that appear when $\Omega=\Omega_{n_y}$, \eqref{bar MM Omega}, by comparing to the analytic solution for the GL growth rate for a black string of length $L$ and radius $r_0=1$ at  large $D$ \cite{Emparan:2013moa},
\begin{equation}
	W_s = \frac{2\pi}{L}\left(1 - \frac{2\pi}{L}\right)\,.
\end{equation}
Close to the marginal mode at $L_0 = 2 \pi$, this behaves as
\begin{equation}
	W_s \approx \frac{L}{2\pi} - 1\, .
\end{equation}

Since the length of a black bar is inversely proportional to its angular velocity, \eqref{longbar}, we relate the relative deviation from the zero modes of the black bar and the string as
\begin{equation}
	\frac{L}{L_0} \approx \frac{\Omega_{n_y}}{\Omega}\, .
\end{equation}
This leads to the estimate
\begin{equation}\label{W estimate}
	W \approx \frac{\Omega_{n_y}}{\Omega} - 1 = \frac{1}{\Omega}\frac{\sqrt{n_y - 1}}{n_y} - 1= \frac{J}{M} \frac{\sqrt{n_y - 1}}{n_y} - 1\, .
\end{equation}

\begin{figure}[th]
\centering
\includegraphics[width=0.6 \linewidth]{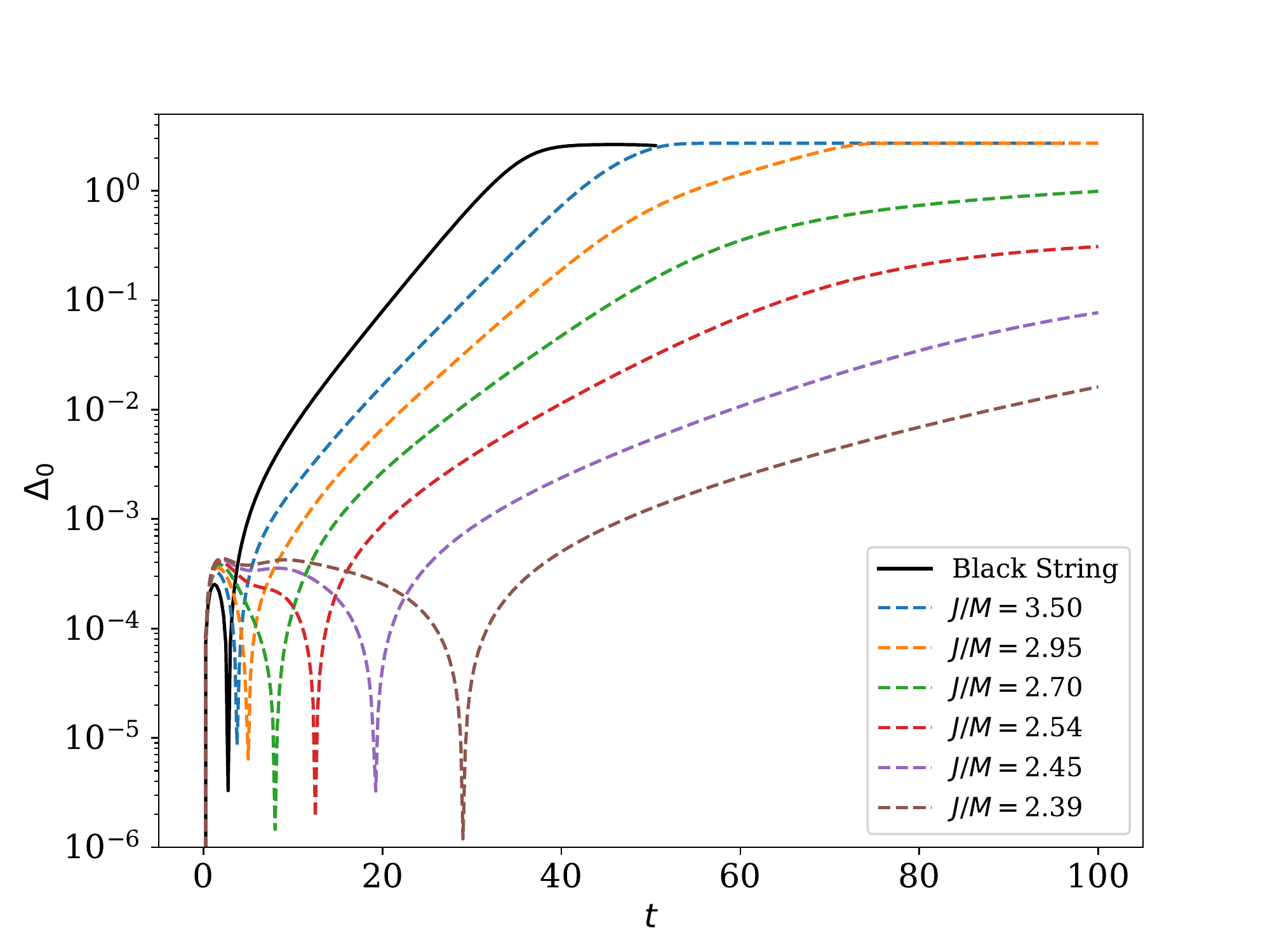}
\caption{\small Deformation of the black bars at their center, as a function of time, for several values of $J/M$. Perturbations of all unstable bars ($J/M > 4/\sqrt{3}$) present a phase of exponential growth given by the linear growth rate $W$ of the dominant mode. The zero mode with $n_y = 4$ is recovered as the growth rate vanishes at the threshold of the instability, $W(J/M = 4/\sqrt{3}) = 0$. Also, for very long bars, $W(J/M \to \infty) \to 1/4$, recovering the dominant growth rate of a black string.}
\label{fig:PerturbedBarEvolution}
\end{figure}

This estimate turns out to be in remarkable agreement with the growth rate of the fundamental symmetric mode $n_y = 4$, as measured from numerical solutions. We extract it from the quantity
\begin{equation}
	\Delta_0 \equiv |m_\text{origin}(t) - m_\text{origin}(t=0)|\,
\end{equation}
\ie the deviation of the central mass density from its initial value. In figure \ref{fig:PerturbedBarEvolution}, $\Delta_0$ is plotted as a function of time for several values of $J/M$. 

\begin{figure}[h]
\centering
\includegraphics[width=0.6 \linewidth]{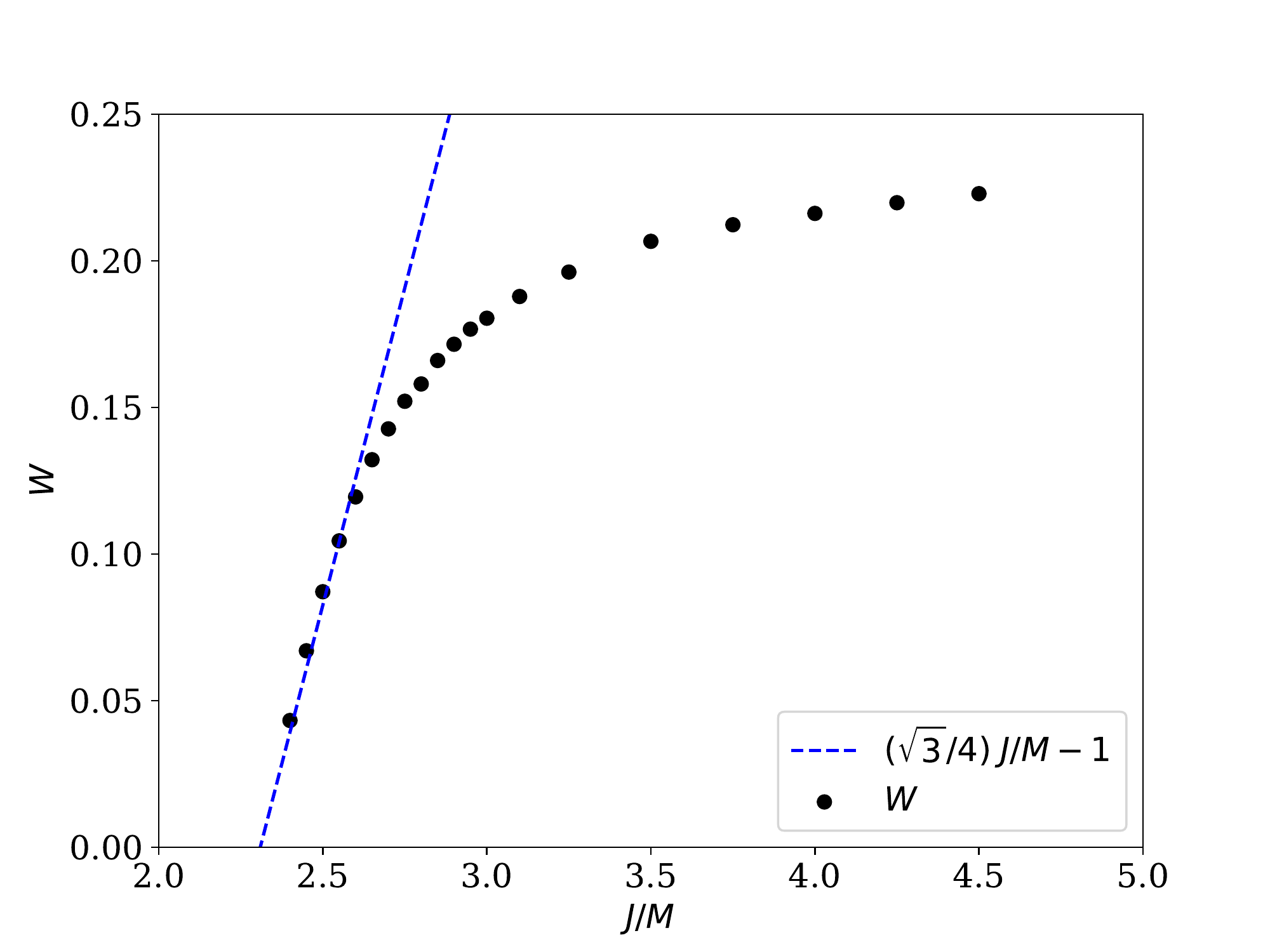}
\caption{\small Growth rate $W$ of the dominant unstable mode of the black bars as a function of $J/M$. The rates are computed by linear regression of the curves in Figure  \ref{fig:PerturbedBarEvolution} at a value of the deformation of $10^{-2}$. The blue dashed line is the black string approximation to a black bar.}
\label{fig:GrowthRatesLine}
\end{figure}

By numerically evaluating the growth rate of the $\Delta_0(t)$ curves in their  phase of exponential growth (in our case, when $\Delta_0 \approx 10^{-2}$), we can obtain an estimate for the instability rate $W$ of the fundamental mode. This is depicted in figure \ref{fig:GrowthRatesLine}, where it is manifest that the black-string-like relation (\ref{W estimate}) is satisfied with remarkable accuracy for values of $J/M$ close to the threshold value for the $n_y = 4$ mode. At large values of $J/M$, $W$ asymptotes to the infinite black string value of $1/4$ as expected.

We conclude that, at least at its onset, the instability of black bars is of the same qualitative and quantitative nature as the GL instability of black strings. This lends further support to the overall picture presented in this article.

\section{Spin-down from gravitational radiation}
\label{sec:grav_rad}

A rotating black bar has a varying quadrupole mass moment and will necessarily emit gravitational radiation in any finite $D$. This radiation will carry away both energy and angular momentum, so the ratio $J/M$ will change over time, possibly decreasing quickly enough that the bar enters a regime of lower $J/M$ where it is stable. If this were the case, instead of proceeding to pinch-off, the bar would spin itself down through gravitational radiation to a stable MP black hole, thus thwarting the evolution towards CC-violation. Already in  \cite{Emparan:2013moa} generic arguments were given that at very large $D$ the emission of radiation is very strongly suppressed; here we attempt to be much more precise about the effect.

Ref.~\cite{Shibata:2010wz}, and more recently \cite{Bantilan:2019bvf}, have followed numerically the evolution of the transient black bars that form from the ultraspinning instability of MP black holes in $D=6,7,8$. Both references find that the black bars return back to MP black holes after radiating their excess spin. Indeed, ref.~\cite{Bantilan:2019bvf} reports huge emissions of the initial mass ($31\%$) and angular momentum ($50\%$) into gravitational waves in $D=6$. 

These results may seem to go against---or at least not provide support for---our claim that black bars become GL-unstable. The simplest interpretation is that in the relatively low dimensions considered in \cite{Shibata:2010wz,Bantilan:2019bvf} the emission is not suppressed enough to let the GL instability grow, but in higher dimensions the latter should dominate. While our arguments make this almost certainly true, we also believe that a stronger case can be made, since we can argue that down to $D=6$, long black bars with large enough spin, if they form, radiate too slowly to prevent the development of the GL instability towards a naked singularity. 

At first this sounds bizarre: shouldn't a long bar with large angular momentum radiate more copiously than one that is shorter and has smaller spin? This reasoning misses the property, seen in \eqref{baromega} and \eqref{longbar}, that black bars with large $J/M$, although very long, rotate very slowly, and this is enough to suppress their wave emission. As we shall show below, the characteristic time for radiative spin-down of a bar of length $\ell_\|$ and angular velocity $\Omega$ is (in units of mass)
\beq\label{tradest}
\tau_\textrm{rad}\sim \frac1{\ell_\|^4\Omega^{D+2}}\,.
\eeq
So, although lengthening the bar accelerates the radiative spin-down, reducing the angular velocity suppresses it---and with a stronger power. For a black bar it is the latter effect that dominates, since $\ell_\|\simeq 1/\Omega$, so
\beq\label{tradest2}
\tau_\textrm{rad}\sim \frac1{\Omega^{D-2}}\,.
\eeq
Hence we expect that slowly rotating, long black bars in any $D\geq 6$ are almost stable to gravitational wave emission. Since they are also GL-unstable, and (as we show below) the growth rate of this instability depends weakly on $D$, we conclude that long black bars die by fragmentation and not by radiation.\footnote{Ref.~\cite{Bantilan:2019bvf} observes lower wave emission at higher spins in four-armed configurations (figure~\ref{fig:QuadrupoleEvolution}), but attributes the effect to the smaller mass in the arms, rather than to slower angular velocity. 
In view of our arguments, the radiative spin down of the black bars observed in \cite{Shibata:2010wz,Bantilan:2019bvf} is the consequence of having too low an initial spin. } However, since shorter bars have been observed to die by radiation in $D=6, 7, 8$, there must be a critical value of the spin per unit mass which separates the two behaviors. This critical value (which currently we cannot compute) will decrease as $D$ grows. 

Then, the question of whether cosmic censorship is violated in a black hole collision in a given dimension $D$ hinges on whether high-spin, supercritical long black bars can form in the merger. To this end, the total spin that can be achieved in the collision is enhanced by having intrinsic spin in the initial MP black holes. This helps, but is limited if we require that these black holes are stable. On the other hand, the initial orbital angular momentum can be increased by enlarging the impact parameter and by increasing the collision velocity. The former is limited by the maximum value for capture. The latter is only limited by the speed of light, but the initial state radiation can grow large as ultrarrelativistic speeds  are approached and may lead to considerable loss of angular momentum. Recall that at large $D$ this radiation is emitted very quickly, in a short burst of duration $\sim 1/D$, and high frequency $\omega \sim D/r_+$.

While we have tried to estimate these effects with all the presently available evidence (including \cite{Yoshino:2002tx,Yoshino:2005hi,Shibata:2008rq,Sperhake:2008ga,Coelho:2012sya,East:2012mb}), we have not been able to reach a definite conclusion for how high a spin can the intermediate state reach in a collision in $D=6, 7$. It seems plausible, though, that in all dimensions $D\gtrsim 8$ collisions can be achieved with high enough total angular momentum such that the intermediate deformed horizon triggers a GL-instability more quickly than radiation spins it down, in accord with the picture that the large-$D$ effective theory has given us.\footnote{We are not invoking decay through four-armed horizons: although at high spins these may grow more quickly than dipole deformations, the structure of the collision strongly favors the development of dipolar bar-like horizons. Nevertheless, four-armed configurations dominate the ultraspinning instability of MP black holes at high spins \cite{Bantilan:2019bvf}. Also, we do not expect spindle bars to drive the evolution: dumbbell bars appear more natural in a collision, and indeed they are the ones we observe in our simulations.} The determination of the actual lower value of $D$ where this is possible will have to await for dedicated numerical simulations in full General Relativity.

\subsection{Radiative spin-down}\label{subsec:TimeScaleOfRad}

Let us now present our estimates of gravitational wave emission from black bars at finite $D$.

We shall model the gravitational wave emission and spin-down of a black bar by assimilating it to a rotating ellipsoid (see appendix \ref{app:barellip}) which radiates according to the $D$-dimensional quadrupole formula. The quadrupolar energy radiation rate in arbitrary (even) $D$ was obtained in \cite{Cardoso:2002pa}. The emission of energy $E$ reduces the mass of the radiating system $\mathbf{M}$ according to $dE=-d\mathbf{M}$. Then (see appendix~\ref{app:energy radn}), for the ellipsoidal bar the relative mass loss rate is 
\begin{align}\label{dEdtbar}
\frac{\dot{\mathbf{M}}}{\mathbf{M}}
&=-\frac{8 (D-3) D }{\pi^{\frac{D-5}{2}} (D-2)(D+1)^{3}\Gamma\left[\frac{D-1}{2}\right]}G \mathbf{M} \mathbf{\Omega}^{D+2}\left(\ell_{\|}^2-\ell_\perp^2\right)^2\nn\\
&=-\frac{\pi D(D-3)^2}{(D-2)(D+1)^{3}}\frac1{\Gamma\lp\frac{D-1}{2}\rp^2}\left(\frac{8G \mathbf{M}}{\Omega_{D-4}} \right)\mathbf{\Omega}^{D+2}\left(\ell_{\|}^2-\ell_\perp^2\right)^2\,.
\end{align}
Here $\mathbf{\Omega}$ is the physical (dimensionful) rotation velocity of the bar. The radiation rate is proportional to $G$ and thus depends on the choice of units. We find convenient to bundle $G\mathbf{M}$ in the term $8G\mathbf{M}/\Omega_{D-4}$,
which is typically proportional (with a coefficient that is weakly dependent on $D$, \ie not exponential nor factorial) to the characteristic horizon radius of the black hole to the power $D-3$. When making comparisons, we will keep this quantity fixed.\footnote{For reference, recall that
$\Omega_{D-4}=\frac{D-3}{2\pi}\Omega_{D-2}=2\pi^{(D-3)/2}/\Gamma\lp\frac{D-3}2\rp$.}

We have also split the numerical $D$-dependent factor in \eqref{dEdtbar} into two terms. The first one  is a rational function that depends weakly on $D$, while the second one yields a factorially suppressed radiation rate at large $D$,
\beq
\frac{\dot{\mathbf{M}}}{\mathbf{M}} \propto -D^{-D} \,.
\eeq

The remaining terms in \eqref{dEdtbar} refer to physical properties of the black bar, namely, its rotation velocity and its shape. They are not independent: for a black bar, in the limit of $D\to\infty$ where it exists as a stationary object, we have 
\beq
\lp\ell_{\|}^2-\ell_{\perp}^2\rp^2=\frac{1-4\left(r_+\mathbf{\Omega}\right)^2}{\mathbf{\Omega}^4}\label{eq:ellminusell}
\eeq
(this follows from \eqref{ls bar} after restoring units, with $r_+$ the horizon radius of the $S^{D-4}$ at the rotation axis). Then 
\begin{align}\label{dMdtbar}
\frac{\dot{\mathbf{M}}}{\mathbf{M}}
&=-\frac{\pi D(D-3)^2}{(D-2)(D+1)^{3}}\frac1{\Gamma\lp\frac{D-1}{2}\rp^2}\lp\frac{8G \mathbf{M}}{\Omega_{D-4}} \rp\mathbf{\Omega}^{D-2}\left(1-4\left(r_+\mathbf{\Omega}\right)^2\right)\,.
\end{align}

When the bar is long, $\ell_{\|}\gg \ell_\perp, r_+$, we find
\begin{align}
\frac{\dot{\mathbf{M}}}{\mathbf{M}} &\propto - \ell_\|^4 \mathbf{\Omega}^{D+2}\\
&\propto - \mathbf{\Omega}^{D-2}
\,.
\end{align}
These results are the basis for the estimate \eqref{tradest}, which is valid in any $D$, and for \eqref{tradest2}, which applies to black bars insofar as \eqref{eq:ellminusell} approximately holds.

In order to obtain the emission rate for a black bar of a given mass and angular velocity, we need $r_+$ in terms of these parameters. Using the leading large-$D$ expression for black bars,
\beq\label{MOmbar}
\mathbf{M}=\frac{\Omega_{D-4}}{8G } \frac{ r_+^{D-4}}{\mathbf{\Omega}}\,,
\eeq
we get
\beq
r_+\mathbf{\Omega}=\lp\frac{8G \mathbf{M}}{\Omega_{D-4}} \rp^{\frac1{D-4}}\mathbf{\Omega}^{\frac{D-3}{D-4}}\,.
\eeq
Notice that at large $D$ and for fixed mass, this is linear in $\mathbf{\Omega}$.

Let us now analyze the dependence on $\mathbf{\Omega}$ in \eqref{dMdtbar}. It vanishes when $ r_+\mathbf{\Omega}=1/2$, which is the bifurcation point with MP black holes; this solution is axisymmetric so it does not radiate. It also vanishes when $\mathbf{\Omega}\to 0$, which is when the bar becomes infinitely long and static. It is maximized at a value of $\mathbf{\Omega}$ slightly below the bifurcation point,
\begin{align}\label{Omrad}
r_+ \mathbf{\Omega}_{\text{max}} = \frac1{2}\lp 1-\frac{c}{D}\rp\,,
\end{align}
where $c$ is a $D$-independent positive number, which we cannot accurately determine without further knowledge of black bars beyond the leading large-$D$ limit.

The maximum radiation rate for a bar of mass $\mathbf{M}$ is then
\begin{align}\label{maxMrate}
\lp\frac{\dot{\mathbf{M}}}{\mathbf{M}}\rp_{\text{max}}
&=-\frac{\pi c}{2^D D}\frac1{\Gamma\lp\frac{D-1}{2}\rp^2}\lp\frac{\Omega_{D-4}}{8G \mathbf{M}} \rp^{\frac{1}{D-3}}\lp 1+\ord{\ln D/D}\rp\,.
\end{align}
We have not expanded the $\Gamma$ functions here since their $D$-dependence even at subleading orders seems robust enough. In any case it is clear that they dominate the large-$D$ behavior.

The radiation rate of angular momentum $\mathcal{ J } $ into gravitational waves in arbitrary (even) $D$ has not been calculated in previous literature. In appendix~\ref{app:ang mom radn} we derive the general quadrupole formula for it. If we then apply it to an object rigidly rotating with angular velocity $\mathbf{\Omega}$ we find the simple relation
\begin{align}\label{dEOdJ}
\dot E =\Omega \dot{\mathcal{J}} \,,
\end{align}
in any dimension $D$.
The spin of the radiating object diminishes as $\dot{\mathbf{J}}=-\dot{\mathcal{ J } }$, and we find
\beq\label{dMOdJ}
\dot{\mathbf{M}}=\mathbf{\Omega} \, \dot{\mathbf{ J }} \,.
\eeq
Our derivation has been made for a rigid, slowly rotating, material solid, but let us apply it to a black hole, and assume that the evolution is slow enough as to proceed along quasistationary configurations. Then the first law of black holes implies that
\beq
\frac{\kappa}{8\pi}\dot{\mathbf{A}}_H=\dot{\mathbf{M}}-\mathbf{\Omega}\dot{\mathbf{ J }}=0\,,
\eeq
namely, the radiation emission is such that entropy production (area increase) is strictly minimized. We stress that this result is independent of $D$, and relies only on the application of the quadrupole emission formula.

For black bars, to leading order at $D\to\infty$, the mass, spin, and angular velocity are related by \eqref{baromega}. In physical magnitudes, using  \eqref{eq:Omegaphys} and \eqref{eq:JMphys}, this  is
\beq
\mathbf{J}=\frac{\mathbf{M}}{(D-2)\mathbf{\Omega}}\,,
\eeq
so \eqref{dMOdJ} gives us
\beq\label{JMrates}
\frac{\dot{\mathbf{J}}}{\mathbf{J}}=D \frac{\dot{\mathbf{M}}}{\mathbf{M}}\lp 1+\ord{\frac1{D}}\rp\,.
\eeq
We see that the relative loss of spin is $D$ times faster than the relative loss of mass, so radiation emission does lead the bar towards smaller values of the spin per unit mass.

We define the characteristic spin-down time as the inverse decay rate of the dimensionless spin per unit mass,
\begin{align}
\tau_{\text{rad}}^{-1}&=-\left(\frac{\mathbf{J}}{\mathbf{M}^{\frac{D-2}{D-3}}}\right)^{-1}\frac{d}{dt}\left(\frac{\mathbf{J}}{\mathbf{M}^{\frac{D-2}{D-3}}}\right)\\
&=-\frac{\dot{\mathbf{J}}}{\mathbf{J}}+\frac{D-2}{D-3}\frac{\dot{\mathbf{M}}}{\mathbf{M}}\,.
\end{align}
For black bars and when $D$ is large, \eqref{JMrates} applies, so
\begin{align}
\tau_{\text{rad}}^{-1}&= -D \frac{\dot{\mathbf{M}}}{\mathbf{M}}\lp 1+\ord{\frac1{D}}\rp\,.
\end{align}
Plugging in \eqref{maxMrate} we obtain our estimate for the fastest radiative spin-down time
\begin{align}\label{fasttaurad}
\tau_{\text{rad}}&=  \frac{2^D }{\pi c}\Gamma\lp\frac{D-1}{2}\rp^2\lp\frac{8G \mathbf{M}}{\Omega_{D-4}} \rp^{\frac{1}{D-3}}\lp 1+\ord{\frac{\ln D}{D}}\rp\,.
\end{align}
The number $c$ is a stand-in for $\ord{1}$ uncertainties from different sources---not only \eqref{Omrad}, but also, \eg $\ord{1}$ corrections to the exponent in $2^D$. These are more important than other uncertainties in our estimates, \eg from properties of black bars at finite $D$ such as \eqref{eq:ellminusell} and \eqref{MOmbar}, which enter only as subleading corrections in $1/D$.
\subsection{Break-up time for the GL-instability}

We estimate the growth rate of the GL instability of the black bar using the results for  a black string in $D$ dimensions obtained in the blackfold approach \cite{Camps:2010br}. This seems to be a very reasonable approximation, given the excellent agreement that we have found in section \ref{sec:BB_evol_inst} for the unstable growth rates of black bars and black strings. The growth time is
\beq\label{tauinst0}
\tau_{\text{inst}}^{-1}=\frac{k}{\sqrt{D-3}}\lp 1-\frac{D-2}{(D-4)\sqrt{D-3}}k r_s\rp\,,
\eeq
where $k$ is the wavenumber of the perturbation and $r_s$ the radius of the black string. This result is valid for all $D$ in an expansion for small $k r_s$. It does not estimate very well the threshold $k$  (where $\tau\to\infty$) at low $D$, but it is close at all $D$ to the known values for the minimum of $\tau_{\text{inst}}$, where the instability is fastest. Since \eqref{tauinst0} is a parabolic profile, the minimum of $\tau_{\text{inst}}$ is at the midpoint
\beq
k_{\text{inst}} r_s=\frac{(D-4)\sqrt{D-3}}{2(D-2)}\,,
\eeq
which gives
\beq\label{tauinst}
\tau_{\text{inst}} =4\frac{D-2}{D-4}r_s\,.
\eeq
Observe crucially that \emph{this does not grow with $D$}, in fact its dependence on $D$ is very weak. It is written in units of $r_s$, but the latter has also a weak dependence on $D$ in units of $8G\mathbf{M}/\Omega_{D-4}$. Indeed, let us translate to mass units. The length of the string that fits this fastest unstable mode is
\beq
L_{\text{inst}}=\frac{2\pi}{k_{\text{inst}}}
\eeq
and the mass of the black string of this length is
\beq
\mathbf{M}=\frac{(D-3)\Omega_{D-3}}{16\pi G}Lr_s^{D-4}\,.
\eeq
Using now that
\beq
\Omega_{D-3}=\sqrt{\pi}\frac{\Gamma\lp\frac{D-3}2\rp}{\Gamma\lp\frac{D-2}2\rp}\Omega_{D-4}\,,
\eeq
we find that, for the fastest unstable black string,
\beq\label{rsinst}
r_s=\lp\frac{8G\mathbf{M}}{\Omega_{D-4}}\frac{D-4}{2\sqrt{\pi}(D-2)\sqrt{D-3}}\frac{\Gamma\lp\frac{D-2}2\rp}{\Gamma\lp\frac{D-3}2\rp}\rp^{\frac1{D-3}}\,.
\eeq
Plugging this into \eqref{tauinst}, we get the GL growth rate for a given mass.
When $8G\mathbf{M}/\Omega_{D-4}$ is fixed, this rate does depend weakly on $D$,
\beq\label{fasttauGL}
\tau_{\text{inst}} =4\lp\frac{8G\mathbf{M}}{\Omega_{D-4}}\rp^{\frac1{D-3}}\lp 1+\ord{\frac{\ln D}{D}}\rp\,.
\eeq

One possible weak link in this estimate is the identification of the instability time $\tau_{\text{inst}}$ with the perturbative result for the growth of the GL instability of a black string. In numerical simulations of the latter, the linear instability starts only after  transients that may last for a few $\tau_{\text{inst}}$, and the time to develop a very large pinch can be significantly larger than $\tau_{\text{inst}}$, possibly even two orders of magnitude larger. Although this would seem to increase the actual value of  $\tau_{\text{inst}}$, on the other hand, as we mentioned earlier, we can expect that the instability of the intermediate black bar formed in a collision proceeds more quickly than the black string instability, mostly because the centrifugal repulsion and the absence of a `box' (the compact Kaluza-Klein direction of the string) will accelerate the development of the instability and push the dumbbell blobs apart. Although we refrain from attempting to estimate these effects, in our simulations at infinite $D$ we do see that the black bars formed in a collision pinch-off more quickly than black strings.

\subsection{Comparing the time scales}

We now have the fastest radiative spin-down time \eqref{fasttaurad} and GL-instability time \eqref{fasttauGL}, both for a given mass in units of $(8G\mathbf{M}/\Omega_{D-4})^{\frac1{D-3}}$. The main finding is almost self-explanatory: 
\beq
\tau_{\text{inst}} =\ord{1} \ll \tau_{\text{rad}}=\ord{D^D}\,.
\eeq
So, for large enough $D$, the radiative spin-down will be so slow as to be negligible. The overall  prefactor in  \eqref{fasttaurad} is in fact larger than the factor $4$ in \eqref{fasttauGL} for all $D\geq 6$ unless $c>9$. Of course these numbers cannot be fully trusted since, \eg the exponent in $2^D$ may easily be, say, $2^{D-3}$, which would make the radiative time faster than the unstable time in $D\approx 6$ with only moderate values of $c$. However, even in this case, having $\tau_{\text{inst}} > \tau_{\text{rad}}$ in $D\gtrsim 8$ would require seemingly unnaturally large values $c\gtrsim 30$, due to the factorial suppression terms.

In conclusion, our estimates suggest that the gravitational emission spin-down will be inefficient to quench the pinch-down instability whenever $D$ is larger than $\approx 8$, and possibly, but much more uncertainly, even down to $D=6$, the lowest dimension where we expect black bars (or similar elongated horizons) to form.

\section{Outlook}
\label{sec:discussion}

Our study of black hole collisions and instabilities admits many potential extensions, \eg to include the effects of rotation in other planes,  corrections in $1/D$, and the inclusion of charge \cite{Andrade:2018rcx}. Although we do not foresee any important qualitative changes to the picture we have presented, each of these extensions also brings in interesting new features, such as entropy production when there is charge or $1/D$ corrections. The introduction of, say, 2-brane charge in the collision plane would, in contrast, hinder the evolution towards pinch-off (localized black holes cannot carry such extended charges).

Without leaving the setup of this article, it should be interesting to perform a more detailed analysis of the $2\to 2$ collisions (and further on, $2\to N$), determining how the final state parameters---the outgoing impact parameter, and the spins and velocity vectors of the outgoing black holes---depend on the initial ones. Our results indicate that, as in conventional scattering theory, when a long-lived, resonant intermediate state forms, it will control the final state. In our case, this is mostly a function of the total $J/M$. But it remains to be investigated in more detail the sensitivity of the final state parameters to initial parameters other than $J/M$.

The extension to collisions between unequal black holes (differing in mass or spin) is readily doable and enlarges the phase space of the collisions. Again, we expect that in much of this phase space the resonant bar with the given value of $J/M$ controls the evolution to the final state, but with possible sensitivity to other particulars.
Likewise, the decay products of the instabilities of ultraspinning MP black holes and black bars admit a much more detailed classification than we have performed here.

These studies should amount to an investigation of the classical scattering of two black holes, and of the decay of unstable states, within the context of a classical theory of blobs, or `lumps' governed by the field equations \eqref{eq brane1}, \eqref{eq brane2}. Although they are non-linear, stationary, localized field configurations, these lumps are not solitons in the strictest sense: their scattering is very non-trivial and is unlikely to be an integrable process. Furthermore, the spectrum of states in the theory is continuous, characterized by two real parameters, the mass and the spin. 
Nevertheless, given the relatively simple nature of the effective equations \eqref{eq brane1}, \eqref{eq brane2}, the scattering perspective in this theory may merit a deeper study, even independently of its interpretation in terms of black holes.

\section*{Acknowledgments}

We are grateful to Pau Figueras for many discussions. Work supported by ERC Advanced Grant GravBHs-692951 and MEC grant FPA2016-76005-C2-2-P. 
RL is supported by Ministerio de Educaci\'on, Cultura y Deporte Grant No. FPU15/01414.
RL gratefully acknowledges the hospitality of the Perimeter Institute for Theoretical Physics, through the Visiting Graduate Fellowship program, where parts of this work were done.
This research was supported in part by
Perimeter Institute for Theoretical Physics. Research at
Perimeter Institute is supported by the Government of
Canada through the Department of Innovation, Science,
and Economic Development, and by the Province of
Ontario through the Ministry of Research and Innovation.

\appendix

\section{Numerical methods}
\label{app:method}

In our numerical studies we have used two independent codes, with equivalent results: one is written in the {\sl Julia} language \cite{juliacode} and the other one in {\sl Mathematica}.
The {\sl Julia} code uses a two-dimensional Fourier grid with FFT differentiation in the spatial directions, and the {\sl DifferentialEquations.jl} package \cite{diffeqns} for time integration.
The {\sl Mathematica} code uses finite differences in the spatial directions and a fourth-order Runge-Kutta method in the time direction.

Note that the equations of motion become singular at $m = 0$. 
In order to handle this singularity, we have implemented two different cutoffs, which yield 
similar results as the cutoff $\epsilon$ is taken to be small. 
First (Method 1), we substitute $m$ by ${\rm max}(m, \epsilon)$ on the right hand side of \eqref{eq brane1}, 
\eqref{eq brane2}, which explicity removes the singularity, and reduces to the original system for $m > \epsilon$.
Second (Method 2), we simply introduce a small constant energy density $\epsilon$ in the initial condition. 
We may say that Method 1 is an `unphysical cutoff', which alters slighty the equations of motion of the system, whereas Method 2 is a `physical cutoff' which does not modify the theory but only the states we consider.

\begin{figure}[th]
\centering
\includegraphics[width=1 \linewidth]{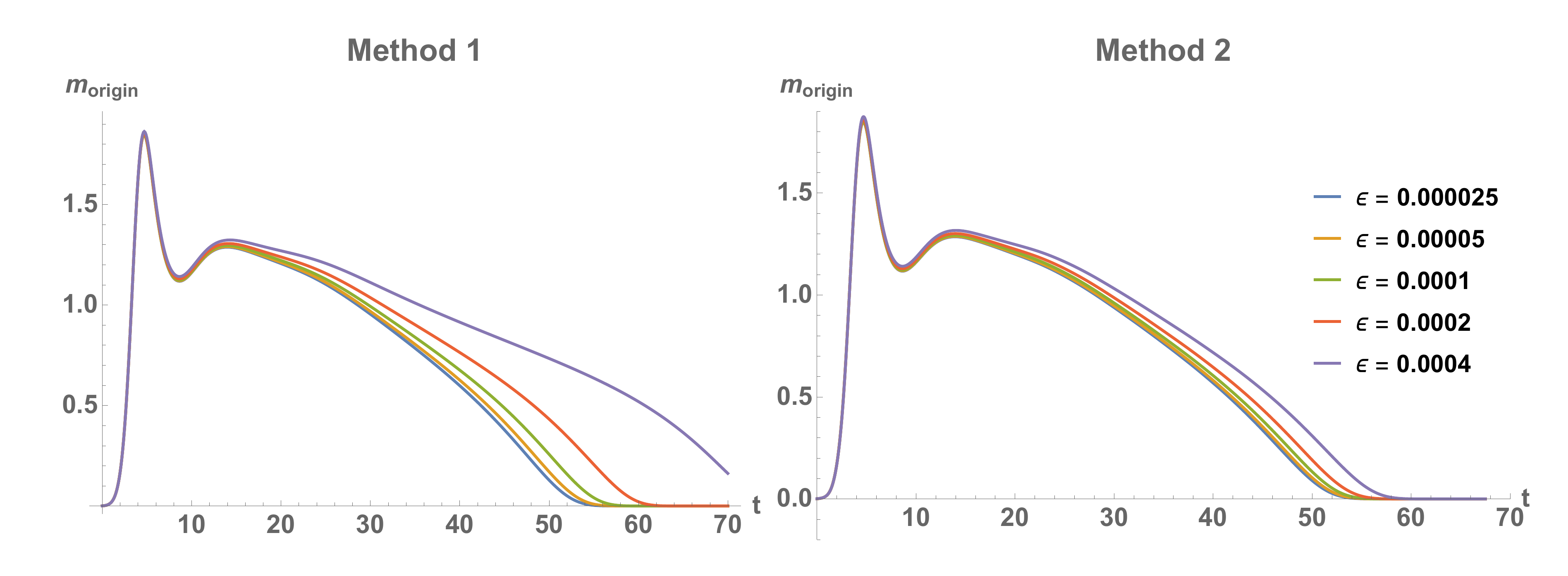}
\caption{\small Dependence of $m_{origin}$ on the low-mass cutoff $\epsilon$, as a function of time, for the two kinds of cutoff introduced in Methods 1 and 2.}%
\label{fig:morigin}
\end{figure}

Let us discuss a concrete example of a collision which yields to CC violation. We take 
$m_0 = 1$, $a = 0.99$, $v = 1.2$, $b = 1.6$, which yields $J/M = 2.94$. 
In figure~\ref{fig:morigin} we show the value of the energy density at 
the origin as a function of time, for various values of the cutoff in both methods. From these results, 
we can read off the break up time, $t_{break}$, defined as the time for which $m_{origin} < 10^{-2}$. 
We see that the dependence with the cutoff is stronger in Method 1, but the behaviour becomes comparable
at small $\epsilon$.

\begin{figure}[th]
\centering
\includegraphics[width=0.5 \linewidth]{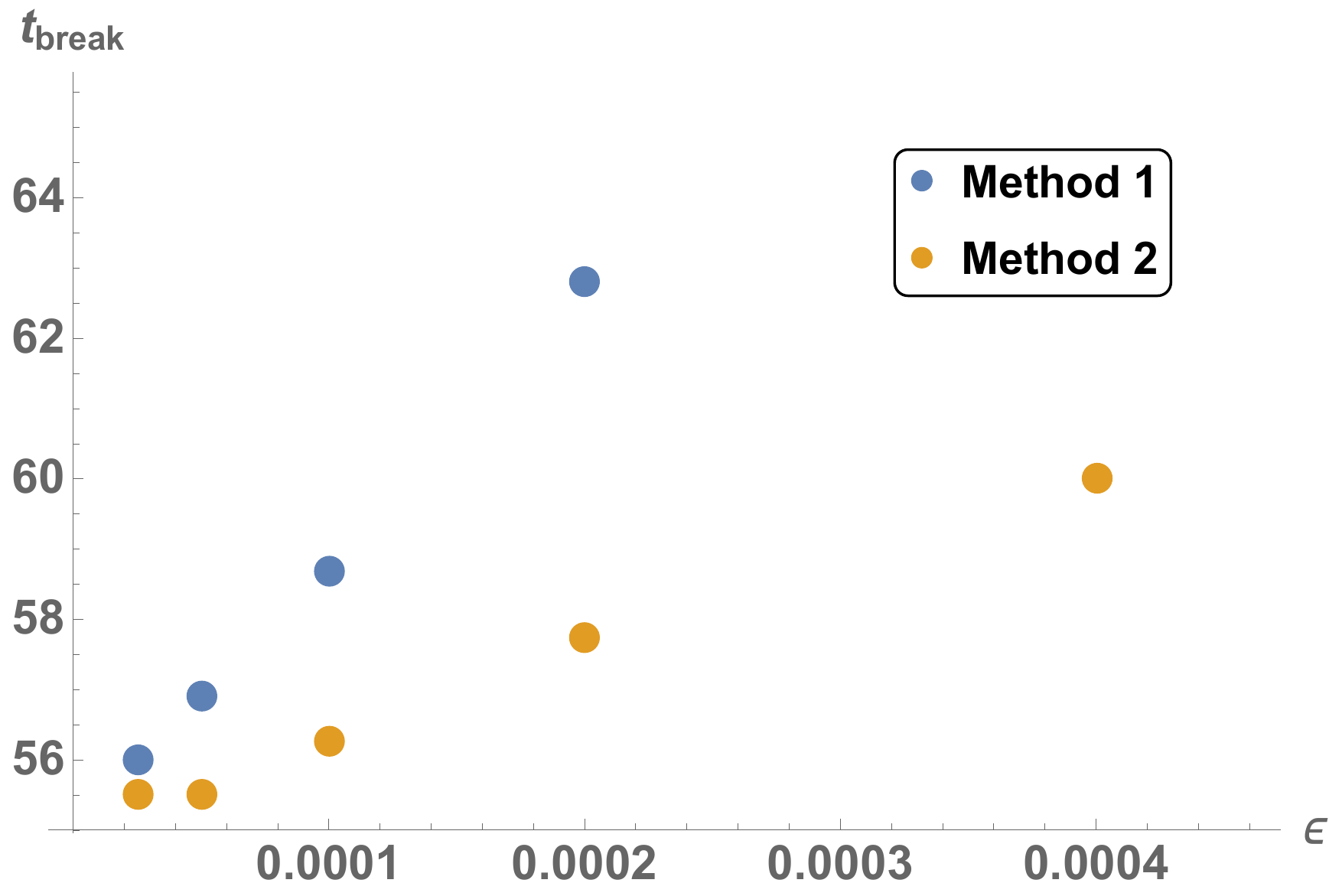}
\caption{\small Dependence of $t_{break}$ as a function of the low-mass cutoff $\epsilon$, for both methods.}%
\label{fig:tbreak_cutoff}
\end{figure}

In figure~\ref{fig:tbreak_cutoff}  we show the dependence of $t_{break}$ with 
$\epsilon$. We see that this dependence is relatively strong, however, there seems to be a 
well-defined limit as we approach $\epsilon = 0$. We expect a closely related behaviour for all 
quantitative results regarding collision final states.


\section{Gravitational radiation in $D$ dimensions}\label{app:grav radn}

We intend to estimate the energy and angular momentum radiated into gravitational waves by a rotating black bar using the quadrupole formula in $D$ dimensions. For this purpose, we begin in appendix~\ref{app:barellip} by modelling a black bar as a rigidly rotating ellipsoid. In appendix~\ref{app:energy radn} we apply to this model the result of \cite{Cardoso:2002pa} for the energy radiation rate. The general formula for angular momentum radiation was not derived in that reference, so we obtain it anew in appendix~\ref{app:ang mom radn}. Then, in appendix~\ref{app:RadRigidRot} we use it to prove that the radiation rates of energy and angular momentum for uniform, rigidly rotating objects (such as the rotating ellipsoidal bar) are related in a simple manner, eq.~\eqref{dEOdJ}.

\subsection{Black bar as a rotating ellipsoid in $D$ dimensions}
\label{app:barellip}
%
We model the black bar by a $D-1$-dimensional spheroid with long axis $\ell_{\|}$, short axis $\ell_\perp$, and radius $r_+$ for the remaining round sphere factor $S^{D-3}$. This is a rigid body described in co-rotating coordinates $y_1(t)$, $y_2(t)$, as
\begin{equation}
\mathfrak{B} = \left\{ y^i \in \mathbb{R}^{D-1} \,\Bigg|\; \frac{y_1^2}{\ell_{||}^2} +\frac{ y_2^2}{\ell_\perp^2} + \frac{1}{r_+^2}\sum_{i=3}^{D-1}y_i^2  \leq 1\right\}\,,
\end{equation}
where $y_1(t)$, $y_2(t)$ are related to the inertial coordinates as in \eqref{corot}. 

Given a mass distribution $T^{00}(t,x)$, we introduce the tensor 
\begin{align}\label{qupole}
M^{ij}&=\int_\mathfrak{B}  d^{D-1}x\, T^{00}(t,x) x^i x^j\,,
\end{align}
from which we subtract the trace to obtain the quadrupole moment tensor
 \begin{equation}
  Q^{ij}=M^{ij}-\frac{1}{D-1}\delta^{ij}M_{kk}\,.
  \label{eq:quadropoledefinition}
 \end{equation}
In order to compute \eqref{qupole} we shall use the integrals,
\begin{align} \int d \Omega _ { D - 2 }\, n _ { i } n _ { j } &= \frac { \Omega _ { D - 2 } } { D - 1 } \delta _ { i j } \label{eq:vectorsphere}\,, \\
  \int d \Omega _ { D - 2 }\, n _ { i } n _ { j } n _ { l } n _ { m } &= \frac {  \Omega _ { D - 2 } } { D ^ { 2 } - 1 } \left( \delta _ { i j } \delta _ { l m } + \delta _ { i l } \delta _ { j m } + \delta _ { i m } \delta _ { j l } \right) \,,
\end{align}
where $n^i=x^i/|\mathbf{x}|$ is the unit vector, and $\Omega_{D-2}$ is the volume of $S^{D-2}$,
\begin{align}
\Omega _ { D - 2 } = \frac { 2 \pi ^ { ( D - 1 ) / 2 } } { \Gamma [ ( D - 1 ) / 2 ] }\,.
\end{align}
Then, assuming a constant mass density $\rho$ for the black bar,
\begin{align}
M^{11}&=\int_{\mathfrak{B}} d^{D-1}y\, T^{00}(0,y) \left(y^1 \cos(\Omega t)-y^2\sin(\Omega t)\right)^2\nn\\
&=\ell_{||}\ell_\perp^3\rho\frac{r_+^{D-3}}{D+1} \frac { \Omega _ { D - 2 } } { D - 1 } \left(\left(\frac{\ell_{||}}{\ell_{\perp}}\right)^2\cos^2(\Omega t)+\sin^2(\Omega t)\right)\,,\nn\\
M^{22}&=\int_{\mathfrak{B}} d^{D-1}y\, T^{00}(0,y) \left(y^1 \sin(\Omega t)+y^2\cos(\Omega t)\right)^2\nn\\
&=\ell_{||}\ell_\perp^3\rho\frac{r_+^{D-3}}{D+1} \frac { \Omega _ { D - 2 } } { D - 1 } \left(\left(\frac{\ell_{||}}{\ell_\perp}\right)^2\sin^2(\Omega t)+\cos^2(\Omega t)\right)\,,\\
M^{12}&=\int_{\mathfrak{B}} d^{D-1}y\, T^{00}(0,y)\left(y^1 \cos(\Omega t)-y^2\sin(\Omega t)\right)\left(y^1 \sin(\Omega t)+y^2\cos(\Omega t)\right)\nn\\
&=\ell_{||}\ell_\perp^3\rho\frac{r_+^{D-3}}{D+1} \frac { \Omega _ { D - 2 } } { D - 1 } \left(\left(\left(\frac{\ell_{||}}{\ell_\perp}\right)^2-1\right)\sin(\Omega t)\cos(\Omega t)\right)\,.
\end{align}
Note that these are the only components that can be time-dependent for our setup.

The mass of the bar is\footnote{In order to unclutter the notation, in this appendix we denote the physical mass and angular velocity by $M$ and $\Omega$, instead of $\mathbf{M}$ and $\mathbf{\Omega}$ as in the main text.}
\begin{align}
M&=\int d^{D-1}x T^{00}(t,x)=\int_{\mathfrak{B}} dy^{D-1} \rho
=\ell_{||}\ell_\perp\rho\frac{r_+^{D-3}}{D-1}\Omega_{D-2}\,.
\end{align}
So we have
\begin{align}
M^{11}&=\frac{M\ell_{\perp}^2}{ D + 1 } \left(\left(\frac{\ell_{||}}{\ell_\perp}\right)^2\cos^2(\Omega t)+\sin^2(\Omega t)\right)\,,\\
M^{22}&=\frac{M\ell_{\perp}^2}{ D + 1 }\left(\left(\frac{\ell_{||}}{\ell_\perp}\right)^2\sin^2(\Omega t)+\cos^2(\Omega t)\right)\,,\\
M^{12}&=\frac{M\ell_{\perp}^2}{ D + 1 }\,, \left(\left(\left(\frac{\ell_{||}}{\ell_\perp}\right)^2-1\right)\sin(\Omega t)\cos(\Omega t)\right)\,.
\end{align}Noting that the trace $M^{ii}$ is time-independent, we can obtain
\begin{align}
(\partial_t)^p Q^{ij}=(\partial_t)^p M^{ij}\,.
\end{align}

\subsection{Radiative power of a black bar}
\label{app:energy radn}

Ref.~\cite{Cardoso:2002pa} obtained the radiative power of a slowly moving distribution of matter in $D$ dimensions, characterized by the tensor $M_{ij}$ \eqref{qupole}, as
\begin{align}\label{eq:radPower}
\frac { d E } { d t } = \frac{2^{2-D} G(D-3) D}{\pi^{\frac{D-5}{2}} \Gamma\left[\frac{D-1}{2}\right]\left(D^{2}-1\right)(D-2)}\left[ ( D - 1 ) \partial _ { t } ^ { \frac{ D +2 }{2} } M _ { i j } ( t ) \partial _ { t } ^ { \frac{ D +2 }{2}} M _ { i j } ( t ) - \left| \partial _ { t } ^ { \frac{ D +2 }{2} } M _ { i i } ( t ) \right| ^ { 2 } \right]\,.
\end{align}
The last term does not contribute for a rotating ellipsoidal bar since we have shown in appendix~\ref{app:barellip} that in this case the trace $M _ { i i }$ is time-independent. Inserting the results for the tensor components we obtain \eqref{dEdtbar}.

\subsection{Angular momentum radiated by slowly moving objects in $D$ dimensions}
\label{app:ang mom radn}

Ref.~\cite{Cardoso:2002pa} did not compute the quadrupolar radiation rate of angular momentum in $D$ dimensions, so we derive it here from the start.

The angular momentum is given by
\begin{equation}
\mathcal{ J } ^ { ij } = \frac { 1 } { 16 \pi  G  } \int d ^ { D-1 } x \left[ - {\delta^ {[i}}_{k}{\delta^ {j]}}_{l} \dot{h} _ { a b } ^ { \mathrm { TT } } x ^ { k } \partial ^ { l } h _ { a b } ^ {  \mathrm { TT } } + 2  {\delta^ {[i}}_{k}{\delta^ {j]}}_{l} h _ { a k } ^ { \mathrm { TT } } \dot { h } _ { a l } ^ { \mathrm { TT } } \right]\,,
\label{eq:totangmom}
\end{equation}
with $h _ { a b } ^ { \mathrm { TT } }$ the gravitational perturbation in transverse traceless gauge. The first summand corresponds to an angular part and the second to the spin part of the spin 2 perturbation. Since we can interpret the integrand of equation (\ref{eq:totangmom}) as the averaged angular momentum density $\langle j^{ij}\rangle$, this implies that the rate of radiated momentum is 
\begin{equation}
\frac{d\mathcal{ J } ^ { ij }}{dt} = \frac { 1 } { 16 \pi  G  } \int d ^ { D-2 } \Omega\, r^{D-2} \left[ - {\delta^ {[i}}_{k}{\delta^ {j]}}_{l} h _ { a b } ^ { \mathrm { TT } } x ^ { k } \partial ^ { l } h _ { a b } ^ {  \mathrm { TT } } + 2  {\delta^ {[i}}_{k}{\delta^ {j]}}_{l} h _ { a k } ^ { \mathrm { TT } } \dot { h } _ { a l } ^ { \mathrm { TT } } \right]\,.
\label{eq:rateofradiatedmom}
\end{equation}
Define $\bar{h}_{\mu\nu}$ via $\bar{h}_{\mu\nu}=h _{\mu\nu}-\frac{1}{2} h \eta_{ \mu \nu }$. In Lorenz gauge $\partial^\mu \bar{h}_{\mu\nu}=0$ it satisfies
\begin{equation}
\square \bar { h } _ { \mu \nu } = - 16 \pi G T _ { \mu \nu }\,.
\end{equation}
Then we have the solution \cite{Cardoso:2002pa}
\begin{equation}
\bar{ h } _ { \mu \nu } ( t , \mathbf{x} ) = - 16 \pi G\int d t ^ { \prime } \int d ^ { D - 1 } \mathbf{x} ^ { \prime } T _ { \mu \nu }\left( t ^ { \prime } , \mathbf{x} ^ { \prime } \right) \mathcal{G}\left( t - t ^ { \prime } , \mathbf{x} - \mathbf{x} ^ { \prime } \right) + \text { homogeneous solutions }\,,
\end{equation}
where we are interested in the retarded Green's function
\begin{equation}
\mathcal{G} ^ { \mathrm { ret } } ( t , \mathbf{x} ) = \frac { 1 } { 4 \pi } \left[ - \frac { \partial } { 2 \pi r \partial r } \right] ^ { ( D - 4 ) / 2 } \left[ \frac { \delta ( t - r ) } { r } \right]\,,
\end{equation}
as long as $D$ is even.
It is convenient to introduce the transverse traceless projector constructed via $P _ { i j } ( \hat { \mathbf { k } } ) = \delta _ { i j } - k _ { i } k _ { j }$,
\begin{equation}
\Lambda _ { i j , k l } ( \hat { \mathbf { k } } ) = P _ { i k } P _ { j l } - \frac { 1 } { D-2 } P _ { i j } P _ { k l }\,.
\end{equation}
Explicitly,
\begin{align}
\Lambda _ { i j , l m } ( \hat { k } ) = &\delta _ { i l } \delta _ { j m } -  \hat { k } _ { j } \hat { k } _ { m } \delta _ { i l } -  \hat { k } _ { i } \hat { k } _ { l } \delta _ { j m }+ \frac { 1 } { D - 2 } \left( - \delta _ { i j } \delta _ { l m } + \hat { k } _ { l } \hat { k } _ { m } \delta _ { i j } + \hat { k } _ { i } \hat { k } _ { j } \delta _ { l m } \right) 
\nn\\&+ \frac { D - 3 } { D - 2 } \hat { k } _ { i } \hat { k } _ { j } \hat { k } _ { l } \hat { k } _ { m }\,.
\label{eq:defLambda}
\end{align}
With this, we can extract the transverse traceless (TT) part of $h _ { k l }$  in harmonic gauge
\begin{equation}
h _ { i j } ^ { \mathrm { TT } } = \Lambda _ { i j , k l } h _ { k l } = \Lambda _ { i j , k l } \bar{h} _ { k l }\,,
\end{equation}
so outside the source we can put the field in TT-gauge
\begin{equation}
h _ { i j } ^ { \mathrm { TT } } ( t , \mathbf{x} ) = 4  G  \Lambda _ { i j , k l } ( \hat { \mathbf { n } } ) \iint dt' d ^ { D-1 } x ^ { \prime } \left[ - \frac { \partial } { 2 \pi r \partial r } \right] ^ { ( D - 4 ) / 2 } \left[ \frac { \delta ( t-t' - |\mathbf{x}-\mathbf{x}'| ) } { |\mathbf{x}-\mathbf{x}'| } \right] T _ { k l } \left( t' , \mathbf{x} ^ { \prime } \right)\,.
\end{equation}
We consider only the part of Green's function with the weakest fall-off since we are interested in an expansion in the wave-zone, which gives
\begin{equation}
h _ { i j } ^ { \mathrm { TT } } ( t , \mathbf{x} ) = -8 \pi  G  \Lambda _ { i j , k l } ( \hat { \mathbf { n } } )  \frac { 1 } { ( 2 \pi r ) ^ { ( D - 2 ) / 2 } } \partial _ { t } ^ {  \frac { D - 4 } { 2 }  } \left[ \int d ^ { D - 1 } \mathbf{x} ^ { \prime } T _ { kl } \left( t - \left| \mathbf{x} - \mathbf{x} ^ { \prime } \right| , \mathbf{x} ^ { \prime } \right) \right]\,.
\end{equation}
Now we expand using the small extent $d$ of the source, defining $r=|\mathbf{ x }|, \hat{\mathbf{ n }}=\frac{\mathbf{ x }}{r}$
\begin{equation}
\left| \mathbf{x} - \mathbf{x} ^ { \prime } \right| = r - \mathbf{x} ^ { \prime } \cdot \hat { \mathbf { n } }  + O \left( \frac { d ^ { 2 } } { r } \right)\,,
\end{equation}
which gives
\begin{equation}
h _ { i j } ^ { \mathrm { TT } } ( t , \mathbf{x} ) = -8 \pi  G  \Lambda _ { i j , k l } ( \hat { \mathbf { n } } ) \frac { 1 } { ( 2 \pi r ) ^ { ( D - 2 ) / 2 } } \partial _ { t } ^ {  \frac { D - 4 } { 2 }  } \left[ \int d ^ { D - 1 } \mathbf{x} ^ { \prime } T _ { kl } \left( t -r - \mathbf{x} ^ { \prime } \cdot \hat { \mathbf { n } } , \mathbf{x} ^ { \prime } \right) \right]\,.
\label{eq:pertinfarfield}
\end{equation}
The next approximation is the Newtonian approximation for slow internal velocities of the source: For this consider the Fourier transform of the stress-energy tensor
\begin{equation}
T _ { k l } \left( t - \frac { r } { c } + \frac { \mathbf{x} ^ { \prime } \cdot \hat { \mathbf { n } } } { c } , \mathbf{x} ^ { \prime } \right) = \int \frac { d ^ { D } k } { ( 2 \pi ) ^ { D } } \tilde { T } _ { k l } ( \omega , \mathbf { k } ) e ^ { - i \omega \left( t - r / c + \mathbf{x} ^ { \prime } \cdot \hat { \mathbf { n } } / c \right) + i \mathbf { k } \cdot \mathbf{x} ^ { \prime } }\,,
\end{equation}
which we can expand in $\omega d$
\begin{equation}
e ^ { - i \omega \left( t - r  + \mathbf{x} ^ { \prime } \cdot \hat { \mathbf { n } }  \right) } = e ^ { - i \omega ( t - r  ) }\left[ 1 - i \omega x ^ { \prime } n ^ { i } + \frac { 1 } { 2 } \left( - i \omega \right) ^ { 2 } x ^ { \prime i } x ^ { \prime j } n ^ { i } n ^ { j } + \ldots \right]\,,
\end{equation}
which corresponds to the expansion in $\mathbf{x}'\cdot\mathbf{\hat{n}}$ ,
\begin{equation}
T _ { k l } \left( t - r + \mathbf { x }' \cdot \hat { \mathbf { n } }  , \mathbf{x} ' \right) \simeq T _ { k l } \left( t - r  , \mathbf { x }' \right)+  x ^ {\prime i } n ^ { i }  \partial _ { 0 } T _ { k l } + \frac { 1 } { 2  } x ^ {\prime i } x ^ {\prime j } n ^ { i } n ^ { j } \partial _ { 0 } ^ { 2 } T _ { k l } + \ldots\,.
\end{equation}
Define the momenta of the stress-energy tensor
\begin{align}
 S ^ { i j } ( t ) & = \int d ^ { D-1 } x T ^ { i j } ( t , \mathbf{x} ) \,,\\ S ^ { i j , k } ( t ) & = \int d ^ { D-1 } x T ^ { i j } ( t , \mathbf{x} ) x ^ { k }\,, \\ S ^ { i j , k l } ( t ) & = \int d ^ { D-1 } x T ^ { i j } ( t , \mathbf{x} ) x ^ { k } x ^ { l }
\,. \end{align}
 With this, eq.~(\ref{eq:pertinfarfield}) becomes
 \begin{align}
 h _ { i j } ^ { \mathrm { TT } } ( t , \mathbf{x} ) & = 8 \pi  G  \Lambda _ { i j , k l } ( \hat { \mathbf { n } } ) \frac { 1 } { ( 2 \pi r ) ^ { ( D - 2 ) / 2 } } \partial _ { t } ^ {  \frac { D - 4 } { 2 }  }\int d ^ { D - 1 } \mathbf{x} ^ { \prime }\int \frac { d ^ { D } k } { ( 2 \pi ) ^ { D } } \tilde { T } _ { k l } ( \omega , \mathbf { k } ) e ^ { - i \omega \left( t - r / c + \mathbf{x} ^ { \prime } \cdot \hat { \mathbf { n } } / c \right) + i \mathbf { k } \cdot \mathbf{x} ^ { \prime } }\nn\\
 & =8 \pi  G  \Lambda _ { i j , k l } ( \hat { \mathbf { n } } ) \frac { (-i\omega)^ { \frac { D - 4 } { 2 }}} { ( 2 \pi r ) ^ { ( D - 2 ) / 2 } } \int d ^ { D - 1 } \mathbf{x} ^ { \prime }\int \frac { d ^ { D } k } { ( 2 \pi ) ^ { D } } \tilde { T } _ { k l } ( \omega , \mathbf { k } ) e ^ { - i \omega \left( t - r / c + \mathbf{x} ^ { \prime } \cdot \hat { \mathbf { n } } / c \right) + i \mathbf { k } \cdot \mathbf{x} ^ { \prime } }\,,
 \end{align}
which we can approximate as
 \begin{align}
 h _ { i j } ^ { \mathrm { TT } } ( \omega, \mathbf{x} ) \simeq 8 \pi  G  \Lambda _ { i j , k l } ( \hat { \mathbf { n } } ) \frac { (-i\omega)^ {  \frac { D - 4 } { 2 }  }} { ( 2 \pi r ) ^ { ( D - 2 ) / 2 } } \tilde{S}_{kl}(\omega)\,.
 \label{eq:farfieldapproximated}
 \end{align}
 Now using conservation of the stress-energy tensor and using \eqref{qupole}
  we can show that
  \begin{equation}
  S ^ { i j } = \frac { 1 } { 2 } \ddot { M } ^ { i j }\,.
  \label{eq:massmomenta}
  \end{equation}
We can write equation (\ref{eq:farfieldapproximated}), using (\ref{eq:massmomenta}), (\ref{eq:quadropoledefinition}), and the properties of the $\Lambda$-tensor
 \begin{align}
   h _ { i j } ^ { \mathrm { TT } } ( \omega, \mathbf{x} ) &\simeq 4\pi  G  \Lambda _ { i j , k l } ( \hat { \mathbf { n } } ) \frac { (-i\omega)^ {  \frac { D - 4 } { 2 }  }} { ( 2 \pi r ) ^ { ( D - 2 ) / 2 } }(-i \omega)^2\tilde{M}_{kl}(\omega)\nn\\
   &= 4\pi  G  \Lambda _ { i j , k l } ( \hat { \mathbf { n } } ) \frac { (-i\omega)^ { \frac { D } { 2 } }} { ( 2 \pi r ) ^ { ( D - 2 ) / 2 } }\tilde{Q}^{kl}(\omega)\nn\\
   &= 4 \pi  G  \frac { (-i\omega)^ { \frac { D } { 2 } }} { ( 2 \pi r ) ^ { ( D - 2 ) / 2 } }\tilde{Q}^{\mathrm{TT}}_{kl}(\omega)\,, \label{eq:Farfieldquadrupoleomega}
  \end{align}
  or equivalently, in position space,
  \begin{align}
   h _ { i j } ^ { \mathrm { TT } } ( t, \mathbf{x} ) &=4 \pi  G  \frac { 1} { ( 2 \pi r ) ^ { ( D - 2 ) / 2 } }\partial_t^{  \frac { D } { 2 } }Q^{\mathrm{TT}}_{kl}(t-r)\,.
 \label{eq:Farfieldquadrupole}
  \end{align}
  Now consider the orbital part of equation (\ref{eq:rateofradiatedmom}) and use (\ref{eq:Farfieldquadrupole})
  \begin{align}
\frac{d\mathcal{ J } ^ { ij }}{dt} &= \frac { 1 } { 16 \pi  G  } \int d ^ { D-2 } \Omega\, r^{D-2} \left[ - {\delta^ {[i}}_{k}{\delta^ {j]}}_{l} \dot{h} _ { a b } ^ { \mathrm { TT } } x ^ { k } \partial ^ { l } h _ { a b } ^ {  \mathrm { TT } }\right]\nn\\
&=\frac{\pi G }{(2\pi)^{D-2}}\int d ^ { D-2 } \Omega \left[ - {\delta^ {[i}}_{k}{\delta^ {j]}}_{l} \partial_t^{  \frac { D+2 } { 2 } }Q _ { a b } ^ { \mathrm { TT } } x ^ { k } \partial ^ { l } \partial_t^{  \frac { D } { 2 } }Q _ { a b } ^ {  \mathrm { TT } }\right]\,.
\label{eq:Angintermsofquadrupole}
  \end{align}
 When calculating $\partial_l Q_{ab}^{\mathrm{TT}}(t-r)=\partial_l\left(\Lambda_{ab,cd}\left( \frac{\mathbf{x}}{r}\right)Q_{cd}(t-r)\right)$ all contributions from derivatives acting on $Q_{cd}$ vanish since $\partial_l Q_{cd}(t-r)=-\frac{x_l}{r}Q_{cd}(t-r)$ which  after antisymmetrization does not  give any contributions to (\ref{eq:Angintermsofquadrupole}). Thus
 \begin{align}
 \frac{d\mathcal{ J } ^ { ij }_{\mathrm{orbit}}}{dt} &=\frac{\pi G }{(2\pi)^{D-2}}\left\langle \partial_t^{  \frac { D+2 } { 2 } }Q _ { cd } \,  \partial_t^{  \frac { D } { 2 } }Q _ { fg }  \right\rangle \int d^{D-2}\Omega \,\Lambda_{ab,cd}x^{[i}\partial^{j]}\Lambda_{ab,fg}\,.
 \end{align}
After some algebra we obtain 
 \begin{equation}
 \frac{d\mathcal{ J } ^ { ij }_{\mathrm{orbit}}}{dt} =\frac{\pi G }{(2\pi)^{D-2}}\left\langle \partial_t^{  \frac { D+2 } { 2 } }Q _ { cd } \,  \partial_t^{  \frac { D } { 2 } }Q _ { fg }  \right\rangle \int d^{D-2} \Omega  \,  {\delta^ {[i}}_{k}{\delta^ {j]}}_{l} n _k \left( n _f \Lambda _ { cd,lg } + n _ { g } \Lambda _ {cd,lf } \right)\,,
 \end{equation}
 which can be calculated using \eqref{eq:defLambda} and  \eqref{eq:vectorsphere}\,.
These integrals are everything we need since the term containing a product of 4 $n_i$ vanishes due to antisymmetrization. With this we obtain
\begin{align}
\frac{d\mathcal{ J } ^ { ij }_{\mathrm{orbit}}}{dt} &=\frac{\pi G }{(2\pi)^{D-2}}\left\langle \partial_t^{  \frac { D+2 } { 2 } }{Q^{[i} }_ { d } \,  \partial_t^{  \frac { D } { 2 } }{Q^{j]}} _ {d}  \right\rangle \frac{2(D-3)D}{(D-2)(D-1)(D+1)}\frac { 2 \pi ^ { ( D - 1 ) / 2 } } { \Gamma [ ( D - 1 ) / 2 ] }\nn\\
&=\frac{ G  (D-3)D}{2^{D-4}\pi^{\frac{D-5}{2}}(D-2)(D-1)(D+1)\Gamma [ ( D - 1 ) / 2 ] } \left\langle \partial_t^{  \frac { D+2 } { 2 } }{Q^{[i} }_ { d } \,  \partial_t^{  \frac { D } { 2 } }{Q^{j]}} _ {d}  \right\rangle\,.
\end{align}
The calculation of the spin part gives
\begin{align}
\frac{d\mathcal{ J } ^ { ij }_{\mathrm{spin}}}{dt} &= \frac { 1 } { 16 \pi  G  } \int d ^ { D-2 } \Omega\, r^{D-2} \left[ 2  {\delta^ {[i}}_{k}{\delta^ {j]}}_{l} h _ { a k } ^ { \mathrm { TT } } \dot { h } _ { a l } ^ { \mathrm { TT } } \right]\nn\\
&=\frac{\pi G }{(2\pi)^{D-2}}\left\langle \partial_t^{  \frac { D+2 } { 2 } }Q _ { mn } \,  \partial_t^{  \frac { D } { 2 } }Q _ { cd }  \right\rangle \int d^{D-2} \Omega  \, 2  {\delta^ {[i}}_{k}{\delta^ {j]}}_{l}\Lambda _ { a l , m n } \Lambda _ { a k , c d }\,.
\end{align}
Using the identity
\begin{equation}
\Lambda _ { a l , m n } \Lambda _ { a k , c d } = P _ { l n } \Lambda _ { m k , c d } - \frac { 1 } { D-2 } P _ { m n } \Lambda _ { k l , c d }\,,
\end{equation}
and antisymmetry we find
\begin{align}
\frac{d\mathcal{ J } ^ { ij }_{\mathrm{spin}}}{dt}&=\frac{2\pi G }{(2\pi)^{D-2}}\left\langle \partial_t^{  \frac { D+2 } { 2 } }Q _ { mn } \,  \partial_t^{  \frac { D } { 2 } }Q _ { cd }  \right\rangle \int d^{D-2} \Omega  \, {\delta^ {[i}}_{k}{\delta^ {j]}}_{l} P _ { l n } \Lambda _ { m k , c d }\nn\\
&=\frac{2\pi G }{(2\pi)^{D-2}}\left\langle \partial_t^{  \frac { D+2 } { 2 } }{Q^{[i} }_ { d } \,  \partial_t^{  \frac { D } { 2 } }{Q^{j]}}_{d}  \right\rangle \frac{(D-3)D}{D^2-1} \frac { 2 \pi ^ { ( D - 1 ) / 2 } } { \Gamma [ ( D - 1 ) / 2 ] }\nn\\
&=\frac{ G (D-3)D}{2^{D-4}\pi^{\frac{D-5}{2}}(D^2-1)\Gamma [ ( D - 1 ) / 2 ]}\left\langle \partial_t^{  \frac { D+2 } { 2 } }{Q^{[i} }_ { d } \,  \partial_t^{  \frac { D } { 2 } }{Q^{j]}}_{d}  \right\rangle\,.
\end{align}
Adding both contributions we obtain the final result
\begin{equation}\label{dJdt}
\frac{d\mathcal{ J } ^ { ij }}{dt}=G\frac{  (D-3)D}{2^{D-4}\pi^{\frac{D-5}{2}}(D-2)(D+1)\Gamma [ ( D - 1 ) / 2 ]}\left\langle \partial_t^{  \frac { D+2 } { 2 } }{Q^{[i} }_ { d } \,  \partial_t^{  \frac { D } { 2 } }{Q^{j]}}_{d}  \right\rangle\,.
\end{equation}

\subsection{Radiation rates of energy and angular momentum for rigidly rotating objects}\label{app:RadRigidRot}

A  body rotating rigidly in the $(1,2)$-plane with an angular velocity $\Omega$, such as the rotating ellipsoidal bar of appendix~\ref{app:barellip}, is described by a mass distribution that will appear static in its inertial frame defined by \eqref{corot}. To simplify calculations it will be convenient to introduce polar coordinates $(\tilde{r},\tilde{\phi})$ for the rotation plane of the inertial frame. These are related to the static coordinates $x^i$ according to
\begin{align}
x^1=y^1\cos(\Omega t)-y^2\sin(\Omega t) = \tilde{r} \cos(\tilde{\phi}+\Omega t)\,,\nn\\
x^2=y^1\sin(\Omega t)+y^2\cos(\Omega t)= \tilde{r} \sin(\tilde{\phi}+\Omega t)\, .
\end{align}
In these coordinates we calculate the momenta of its mass distribution
\begin{align}
M^{11}&=\int d^{D-1}y\, T^{00}(0,y) \left(\tilde{r} \cos(\tilde{\phi}+\Omega t) \right)^2\nn\\
&=\int d^{D-1}y\, T^{00}(0,y) \frac{\tilde{r}^2}{2} \left(1+\cos(2\tilde{\phi})\cos(2\Omega t)-\sin(2\tilde{\phi})\sin(2\Omega t) \right)\nn\\
&= \widetilde{M}^{11} \cos(2\Omega t) -\widetilde{M}^{22}\sin(2\Omega t) +\text{const.}\,,\\
M^{22}&=\int d^{D-1}y\, T^{00}(0,y) \left(\tilde{r} \sin(\tilde{\phi}+\Omega t) \right)^2\nn\\
&= -\widetilde{M}^{11} \cos(2\Omega t) +\widetilde{M}^{22}\sin(2\Omega t) +\text{const.}\,,\\
M^{12}&=\int d^{D-1}y\, T^{00}(0,y)\left(\tilde{r}^2 \cos(\tilde{\phi}+\Omega t)\sin(\tilde{\phi}+\Omega t)\right)\nn\\
&= \widetilde{M}^{11} \sin(2\Omega t) +\widetilde{M}^{22}\cos(2\Omega t) +\text{const.}\,,
\end{align}
where we defined 
\begin{align}
\widetilde{M}^{11}=\frac12 \int d^{D-1}y \,T^{00}(0,y) y^1y^1\,,\nn\\
\widetilde{M}^{22}=\frac12 \int d^{D-1}y \,T^{00}(0,y) y^2y^2\,,
\end{align}
 as the momenta of the mass distribution in the co-rotating frame.
 
 With this we can evaluate the tensor structures appearing in eq.~\eqref{eq:radPower}, \eqref{dJdt}. Noting again that the trace does not show time dependence we calculate
 \begin{align}
\partial_{t}^{\frac{D+2}{2}} M_{i j}(t) \partial_{t}^{\frac{D+2}{2}} M_{i j}(t)=2(2\Omega)^D  \left( (\widetilde{M}^{11})^2 +(\widetilde{M}^{22})^2\right)\,,
 \end{align}
and
\begin{align}
\partial_t^{  \frac { D+2 } { 2 } }{M^{[1} }_ { d } \,  \partial_t^{  \frac { D } { 2 } }{M^{2]}}_{d} =\frac{1}{2\Omega} (2\Omega)^D\left( (\widetilde{M}^{11})^2 +(\widetilde{M}^{22})^2\right)\,.
\end{align}
Inserting these into eq.~\eqref{eq:radPower} and \eqref{dJdt}, and comparing the results, we obtain \eqref{dEOdJ}.

\end{document}